\pdfoutput = 1
\documentclass[journal=nalefd, manuscript=letter]{achemso}
\usepackage{graphicx}
\usepackage{graphics}
\usepackage{amsmath}
\usepackage{amssymb}
\usepackage{amsthm}
\usepackage{amsfonts}
\usepackage{tikz}
\usepackage{dcolumn}
\usepackage{dsfont}
\usepackage{latexsym}
\usepackage{rotating}
\usepackage{array}
\usepackage{makecell}
\usepackage{color}
\usepackage{bbm}
\usepackage{physics}
\usepackage{subfigure}
\usepackage{float}
\usepackage{epsfig}
\usepackage{epsf}
\usepackage{psfrag}
\usepackage{bm}
\usepackage{mathrsfs}
\usepackage{tabularx}
\usepackage{url}
\usepackage{epstopdf}
\usepackage{xcolor}
\usepackage{setspace}

\newcommand{\onlinecite}[1]{\hspace{-1 ex} \nocite{#1}\citenum{#1}}
\newcommand{\lllangle}{\langle\!\langle\!\langle}
\newcommand{\rrrangle}{\rangle\!\rangle\!\rangle}
\newcommand{\unitvec}[1]{\hat{\mathbf{#1}}}			% Unit vectors with hat
\newcommand{\ii}{\mathrm{i}}

\usepackage{color} % use colored text in latex

\usepackage{hyperref}
\hypersetup{
colorlinks=true,
        final=true,
        linkcolor=black,
        citecolor=black,
        filecolor=black,
        urlcolor=black,
}

\title{
\Large
\setstretch{1}
Tuning the magnetic properties of Kitaev materials via the antiferromagnetic proximity effect: Novel phases and application to an $\alpha$-RuCl$_3$/MnPS$_3$ bilayer
}

\author{Pedro M. C\^onsoli}
\affiliation{Department of Physics, Arizona State University, Tempe, AZ 85287, USA}
\email{pconsoli@asu.edu}

\author{Ezra Day-Roberts}
\affiliation{Department of Physics, Arizona State University, Tempe, AZ 85287, USA}

\author{Johannes Knolle}
\affiliation{Munich Center for Quantum Science and Technology (MCQST), Schellingstrasse 4, 80799 Munich, Germany}
\alsoaffiliation{Technical University of Munich, TUM School of Natural Sciences, Physics Department, TQM, 85748 Garching,
Germany}
\alsoaffiliation{Blackett Laboratory, Imperial College London, London SW7 2AZ, United Kingdom}

\author{Antia S. Botana}
\affiliation{Department of Physics, Arizona State University, Tempe, AZ 85287, USA}
\email{antia.botana@asu.edu}
\author{Onur Erten}
\affiliation{Department of Physics, Arizona State University, Tempe, AZ 85287, USA}
\email{onur.erten@asu.edu}

\begin{document}

\setstretch{1.2}


%%%%%%%%%%%%%%%%%%%%%%%%%%%%%%%%%%%%%%%%%%%%%%%%%%%%%%%%%%%%%%%%%%%%%%%%%%%%%%%
% Abstract
%%%%%%%%%%%%%%%%%%%%%%%%%%%%%%%%%%%%%%%%%%%%%%%%%%%%%%%%%%%%%%%%%%%%%%%%%%%%%%%

\begin{abstract}

In recent years, the increasing level of control over van der Waals (vdW) heterostructures has opened new routes to tune the properties of quantum materials. Motivated by these developments, we examine the potential consequences of interfacing a Kitaev honeycomb magnet, such as $\alpha$-RuCl$_3$, with a nearly lattice-matched vdW antiferromagnet.
By combining perturbation theory, exact diagonalization, and a classical energy-minimization method, we show that an effective staggered magnetic field originating from the vdW antiferromagnet can drive a monolayer of a Kitaev material into various novel phases, including an antichiral Kitaev spin liquid, a nonmagnetic nematic phase, and different types of skyrmion crystals.
We then apply first-principles simulations to assess the prospect of concretely realizing this setup in a heterobilayer of $\alpha$-RuCl$_3$ and the easy-axis antiferromagnet MnPS$_3$.

\end{abstract}

%%%%%%%%%%%%%%%%%%%%%%%%%%%%%%%%%%%%%%%%%%%%%%%%%%%%%%%%%%%%%%%%%%%%%%%%%%%%%%%
% Introduction
%%%%%%%%%%%%%%%%%%%%%%%%%%%%%%%%%%%%%%%%%%%%%%%%%%%%%%%%%%%%%%%%%%%%%%%%%%%%%%%

A long-standing quest that has united the fields of condensed matter physics and materials chemistry for decades is the search for experimental realizations of quantum spin liquids (QSLs) -- elusive states of matter that display fascinating properties such as long-range entanglement, fractionalized excitations, and emergent gauge fields\cite{Savary2016, Zhou_RMP2017, Knolle_ARCMP2019, Senthil_Science2020}.
By now, various classes of Mott insulators have been identified as candidates to host QSLs, including the so-called Kitaev materials\cite{Trebst_review}. At low energies, these compounds develop spin-orbit entangled $J_\mathrm{eff} = 1/2$ magnetic moments which interact primarily through the bond-dependent Ising exchange couplings observed in Kitaev's honeycomb model\cite{kitaev2006anyons, Jackeli_PRL2009, Chaloupka_PRL2010}. Yet while the latter is celebrated for harboring an exact QSL ground state, Kitaev materials tend to exhibit long-range magnetic order at low temperatures due to subleading magnetic interactions~\cite{Chaloupka_PRL2010, Chaloupka_PRL2013, Rau_PRL2014, Winter_PRB2016, winter2017_review, rousochatzakis2023_review}.

In this context, there have been continued efforts to understand whether external perturbations can stabilize proximate QSLs as ground states of Kitaev magnets. Although this strategy has not been conclusively successful so far, it has generated tremendous activity, especially with regard to $\alpha$-RuCl$_3$. In this vdW material, magnetic Ru$^{3+}$ ions form layers of honeycomb lattices\cite{plumb_PRB2014, johnson_PRB2015} and antiferromagnetic (AFM) zigzag order emerges below a temperature $T_\mathrm{N} \sim 7$~K\cite{Sears_PRB2015, Cao_PRB2016, Banerjee_NatMat2016, Banerjee_Science2017, Ojeda_arXiv2025}.
However, a moderate in-plane magnetic field suppresses the zigzag phase \cite{johnson_PRB2015, Sears_PRB2015, Sears_PRB2017} and drives $\alpha$-RuCl$_3$ into a disordered intermediate-field regime whose nature remains under debate to this day\cite{Baek_PRL2017,Wolter_PRB2017,Kasahara_Nature2018,Yokoi_Science2021,Gass_PRB2020,lefranccois2022,Bruin_NatPhys2022,czajka_NatMat2023,Sarkis2026}.

In parallel, various other routes have been explored as means of uncovering new phases in $\alpha$-RuCl$_3$. Apart from chemical substitution, carrier doping~\cite{bastien_PRB2019}, strain engineering~\cite{bastien2018,bachus2020,kaib2021,kocsis2022} and twisting\cite{Akram_NanoLett2024}, an appealing alternative is to interface $\alpha$-RuCl$_3$ with a different vdW material.
A first example along these lines is provided by $\alpha$-RuCl$_3$/graphene heterostructures, where charge transfer and screening can lead to novel correlated states at the interface\cite{mashhadi2019, zhou2019, biswas2019, Rizzo_NanoLett2020, gerber2020, leeb_PRL2021, balgley2022, shi2023magnetic}. 
A second example involves $\alpha$-RuCl$_3$/CrX$_3$ (X= I,Cl) heterostructures, in which the ferromagnetic (FM) order in CrX$_3$ acts on $\alpha$-RuCl$_3$ as an effective magnetic field with a dominant \emph{uniform} spatial component\cite{Zhang_PRB2024}. 

Here, we examine a related setting where a Kitaev material is subject to a \emph{staggered} magnetic field, which may originate from an adjacent vdW material with AFM order.
A primary motivation to pursue this route is that Kitaev models with FM and AFM couplings have starkly different responses to external magnetic fields: whereas the AFM model exhibits richer physics in a spatially uniform field, the converse occurs in a staggered field\cite{Hickey_NatComm2019}.
Since a significant fraction of Kitaev materials -- including $\alpha$-RuCl$_3$ and other compounds whose magnetic ions have $d^5$ or $d^7$ electronic configurations\cite{Jackeli_PRL2009, Chaloupka_PRL2010,Liu_PRB2018, Sano_PRB2018} -- are believed to possess FM Kitaev couplings, it is natural to ask what new phases they might realize in the proposed AFM proximity environment.
We tackle this question by studying the influence of a staggered magnetic field on two-dimensional spin models that interpolate between a pure FM Kitaev model and parameter sets previously proposed for $\alpha$-RuCl$_3$. 
As this setup can potentially be realized in practice by interfacing a monolayer of a Kitaev magnet with a vdW antiferromagnet, we also perform first-principles simulations in a heterobilayer of $\alpha$-RuCl$_3$ and the easy-axis antiferromagnet MnPS$_3$.
We find that when a weak staggered field is applied to a Kitaev spin liquid (KSL), it gives rise to Majorana Fermi surfaces\cite{Takikawa_PRB2019,Nakazawa_PRB2022,Soares_arXiv2026} and antichiral edge states\cite{Colomes_PRL2018}. At higher fields, a nematic phase without long-range magnetic order may emerge. In contrast, when the staggered field is applied to a zigzag phase, it induces a metamagnetic transition into at least one novel and robust state with long-range magnetic order. We propose that this state is a skyrmion crystal and discuss the possibility of it being realized in an $\alpha$-RuCl$_3$/MnPS$_3$ bilayer.

%%%%%%%%%%%%%%%%%%%%%%%%%%%%%%%%%%%%%%%%%%%%%%%%%%%%%%%%%%%%%%%%%%%%%%%%%%%%%%%
% Model
%%%%%%%%%%%%%%%%%%%%%%%%%%%%%%%%%%%%%%%%%%%%%%%%%%%%%%%%%%%%%%%%%%%%%%%%%%%%%%%

The starting point for our model calculations is an effective Hamiltonian
\begin{equation}
    \mathcal{H} = \mathcal{H}_K + g\mathcal{H}_{J\Gamma\Gamma'} + \mathcal{H}_h
    \label{eq:H}
\end{equation}
describing a single honeycomb layer of (pseudo)spin-$1/2$ moments $\mathbf{S}_i$ in a Kitaev material. Although the model does not include additional degrees of freedom, it will be instructive to refer to the aforementioned setup, in which the Kitaev layer ($L_1$) is interfaced with a monolayer ($L_2$) of a vdW magnet whose spins $\mathbf{s}_i$ display collinear Néel order. The three terms in Eq.~\eqref{eq:H} read:
\begin{align}
    \mathcal{H}_K &= K_1 \sum_{\gamma=x,y,z} \sum_{\langle ij\rangle_\gamma} S_i^\gamma S_j^\gamma,
    \notag \\
    \mathcal{H}_{J\Gamma\Gamma'} &= 
    J_1 \sum_{\langle ij\rangle} \mathbf{S}_i \cdot \mathbf{S}_j + J_3\sum_{\lllangle ij \rrrangle} \mathbf{S}_i \cdot \mathbf{S}_j
    \notag \\
    &+ \sum_{\gamma} \sum_{\langle ij \rangle_\gamma} \left[ 
    \Gamma_1 \left( S_i^\alpha S_j^\beta + S_i^\beta S_j^\alpha \right)
    + \Gamma_1' \left( S_i^\gamma S_j^\alpha + S_i^\gamma S_j^\beta + S_i^\alpha S_j^\gamma + S_i^\beta S_j^\gamma \right)
    \right],
    \notag \\
    \mathcal{H}_h &= - \sum_i \mathbf{h}_i \cdot \mathbf{S}_i
    \approx -\sum_i \left(J_\perp \mathbf{s}_i\right)\cdot \mathbf{S}_i.
    \label{eq:Hparts}
\end{align}
The first part, $\mathcal{H}_K$, is a Kitaev Hamiltonian with isotropic couplings along the three types of nearest-neighbor (NN) bonds, $\gamma = x,y,z$.
The second term, $\mathcal{H}_{J\Gamma\Gamma'}$, includes other symmetry-allowed interactions that are relevant to Kitaev magnets; they consist of NN and third-NN Heisenberg exchanges, $J_1$ and $J_3$, as well as two off-diagonal, bond-dependent couplings, $\Gamma_1$ and $\Gamma_1'$. For a given NN $\gamma$ bond, the latter are defined in terms of a permutation $(\alpha,\beta,\gamma)$ of $(x,y,z)$.
Finally, the last part represents the influence of an AFM proximity environment by a Zeeman field $\mathbf{h}_i$. As advertised, we will assume that this field has a staggered structure $\mathbf{h}_i = (-1)^i h\unitvec{c}$, with $(-1)^i = \pm1$ on the $A$ ($B$) sublattice of the honeycomb lattice.
We expect $\mathcal{H}_h$ to reasonably capture the low-temperature properties of the alluded $L_1/L_2$ heterostructure if (i) the layers are nearly lattice-matched and (ii) the Néel temperature of $L_2$ is much higher than that of $L_1$. These conditions are met by MnPS$_3$, which has a Néel temperature  $T_\mathrm{N} = 78$~K\cite{Kurosawa_JPSJ1983,Wildes_JPCM1998}, and together allow one to relate the Zeeman term to an interlayer coupling $J_\perp$; see Eq.~\eqref{eq:Hparts}.
For concreteness, we fix the direction of the field to $\unitvec{c} = \frac{1}{\sqrt{3}} \left( \unitvec{x} + \unitvec{y} + \unitvec{z}\right)$. This coincides with the vector normal to the honeycomb planes in real Kitaev materials\cite{Trebst_review,Janssen_JPCM2019} and is consistent with the direction of the easy axis of MnPS$_3$ \cite{Han_PRB2023}.

To reduce the number of free parameters in Eq.~\eqref{eq:H}, we focus on the two sets of couplings $(K_1,J_1,\Gamma_1,\Gamma_1',J_3)$ listed in Table~\ref{tab:edmodels}. The first set was proposed based on a comparison of \textit{ab initio} results and neutron scattering data for $\alpha$-RuCl$_3$,\cite{Winter_NatComm2017} whereas the second followed from a comprehensive analysis of different experiments\cite{Moeller_PRB2025}. As a result, we are left with two free parameters: $h$ and $g$, which we will use to interpolate between a pure Kitaev Hamiltonian ($g=0$) and a more realistic model for $\alpha$-RuCl$_3$ ($g=1$). The choice of examining different models enables us to identify phases that are robust to reasonable variations in the subleading magnetic interactions. We now proceed to describe such phases, placing particular emphasis on the two most interesting instances: the QSL obtained at small $g,h\ne0$ and a magnetically ordered state that emerges once the zigzag order at large $g$ is suppressed by $h$.

\begin{table}[tb!]
\caption{Parameter sets utilized in our ED calculations. Both of these models were previously employed to reproduce experimental results for $\alpha$-RuCl$_3$. The couplings are given in units of meV.}
    \centering
    \begin{tabularx}{\linewidth}{>{\centering\arraybackslash}X >{\centering\arraybackslash}X >{\centering\arraybackslash}X >{\centering\arraybackslash}X>{\centering\arraybackslash}X >{\centering\arraybackslash}X}
    \hline 
    \hline
        Model & $K_1$ & $J_1$ & $\Gamma_1$ & $\Gamma_1'$ & $J_3$ \\
        \hline
        1 [Ref.~\onlinecite{Winter_NatComm2017}] & $-5.0$ & $-0.5$ & 2.5 & 0.0 & 0.5 \\
        2 [Ref.~\onlinecite{Moeller_PRB2025}] & $-7.567$ & $-4.750$ & 4.276 & 2.326 & 3.400 \\
        \hline
        \hline
    \end{tabularx}
    \label{tab:edmodels}
\end{table}

%%%%%%%%%%%%%%%%%%%%%%%%%%%%%%%%%%%%%%%%%%%%%%%%%%%%%%%%%%%%%%%%%%%%%%%%%%%%%%%
% Perturbation theory
%%%%%%%%%%%%%%%%%%%%%%%%%%%%%%%%%%%%%%%%%%%%%%%%%%%%%%%%%%%%%%%%%%%%%%%%%%%%%%%

\begin{figure}[t]
	\center
	\includegraphics[width=\textwidth]{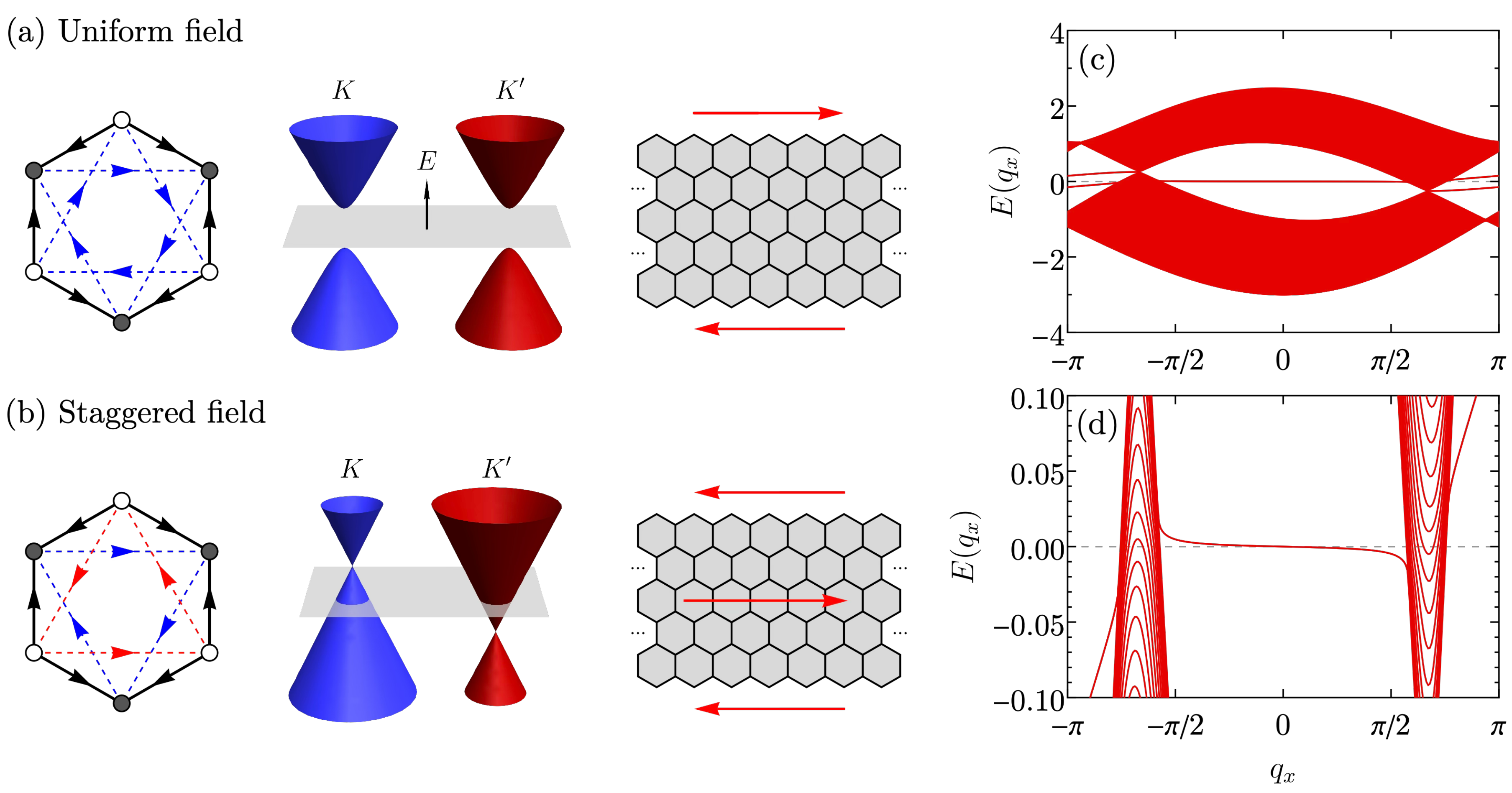} 
	\caption{
		Results obtained from third-order perturbation theory in $h$ around the exactly-solvable Kitaev point. (a,b) Qualitative differences between the cases of a uniform and staggered external magnetic field.
		Left: Next-NN Majorana hopping terms induced by $h$. The dashed arrows are correlated with the sign of the complex hopping amplitudes (see SI). 
		Middle: Majorana dispersion near the two inequivalent corners, $\bm{K}$ and $\bm{K}'$, of the first Brillouin zone. The gray plane corresponds to zero energy.
		Right: Illustration of (a) chiral and (b) antichiral edges modes on a ribbon with zigzag edges.
		(c) Spectrum of the effective Majorana model for a system placed on a ribbon with zigzag edges and subject to a weak staggered field. $q_x$ denotes the momentum along the direction subject to periodic boundary conditions. (d) Zoom at low energies, showing the presence of Majorana Fermi surfaces and gapless edge modes. The parameters used in the calculations are specified in the SI.
	}
	\label{fig:pt}
\end{figure}

Given that the KSL has a nonzero spin gap\cite{kitaev2006anyons}, the system should remain in a QSL phase for small $g,h\ne0$. From a fundamental standpoint, this regime is particularly compelling because its low-energy excitations are emergent Majorana fermions -- charge-neutral fermionic particles that, unlike neutrons and neutrinos, are their own antiparticles and do not exist in isolation. Beyond purely fundamental interest, the realization of these exotic excitations has been actively pursued due to their potential applications in topological quantum computation\cite{Alicea_RPP2012}.
To investigate the nature of the QSL at $h\ne0$ and connect with previous work\cite{kitaev2006anyons,Hickey_NatComm2019,Takikawa_PRB2019,Nakazawa_PRB2022}, we perform a perturbative calculation in $h$ at $g=0$. As detailed in the Supporting Information (SI), this maps the Hamiltonian to an effective hopping problem of Majorana fermions on the honeycomb lattice. Crucially, because the next-NN hopping processes are sensitive to the sign structure of $\mathbf{h}_i$ [see Figs.~\ref{fig:pt}(a,b)], the Majorana fermions acquire \emph{qualitatively} different dispersions in a uniform and staggered field.
On the one hand, a uniform field gaps out the pair of Dirac cones that are pinned to zero energy for $h=0$\cite{kitaev2006anyons}. On the other hand, a staggered field shifts those Dirac cones by opposite energies and generates Majorana Fermi surfaces\cite{Takikawa_PRB2019,Chari_PRB2021,Nakazawa_PRB2022,Soares_arXiv2026}, i.e. analogs of electronic Fermi surfaces for Majorana bands.

The existence of such Majorana Fermi surfaces is evident in the numerical results of Figs.~\ref{fig:pt}(c,d), which display the spectrum of the same effective Majorana model derived from perturbation theory, but now for a semi-infinite system with parallel zigzag edges (see SI). Zooming in at low energies [Fig.~\ref{fig:pt}(d)], we find additional (twofold degenerate) modes that cross zero energy with a negative group velocity at momentum $q_x=0$. Since these modes only exist in the presence of open boundaries, they represent copropagating edge states, which, in equilibrium, must be offset by a counterpropagating current in the gapless bulk. Similar ``antichiral'' edge states were reported in Ref.~\onlinecite{Colomes_PRL2018} for a variant of the Haldane model\cite{Haldane_PRL1988} with the same symmetries as our effective Majorana model.
Altogether, these results highlight that the antichiral KSL found in our models at $h\ne 0$ is fundamentally different from the chiral spin liquid obtained in a weak uniform field\cite{kitaev2006anyons,Hickey_NatComm2019}. This intriguing antichiral KSL should leave distinctive signatures in spatially resolved \emph{thermal} transport measurements, since Majorana fermions carry heat\cite{Kasahara_Nature2018}.

%%%%%%%%%%%%%%%%%%%%%%%%%%%%%%%%%%%%%%%%%%%%%%%%%%%%%%%%%%%%%%%%%%%%%%%%%%%%%%%
% Exact diagonalization
%%%%%%%%%%%%%%%%%%%%%%%%%%%%%%%%%%%%%%%%%%%%%%%%%%%%%%%%%%%%%%%%%%%%%%%%%%%%%%%

\begin{figure}[t]
    \center
    \includegraphics[width=\textwidth]{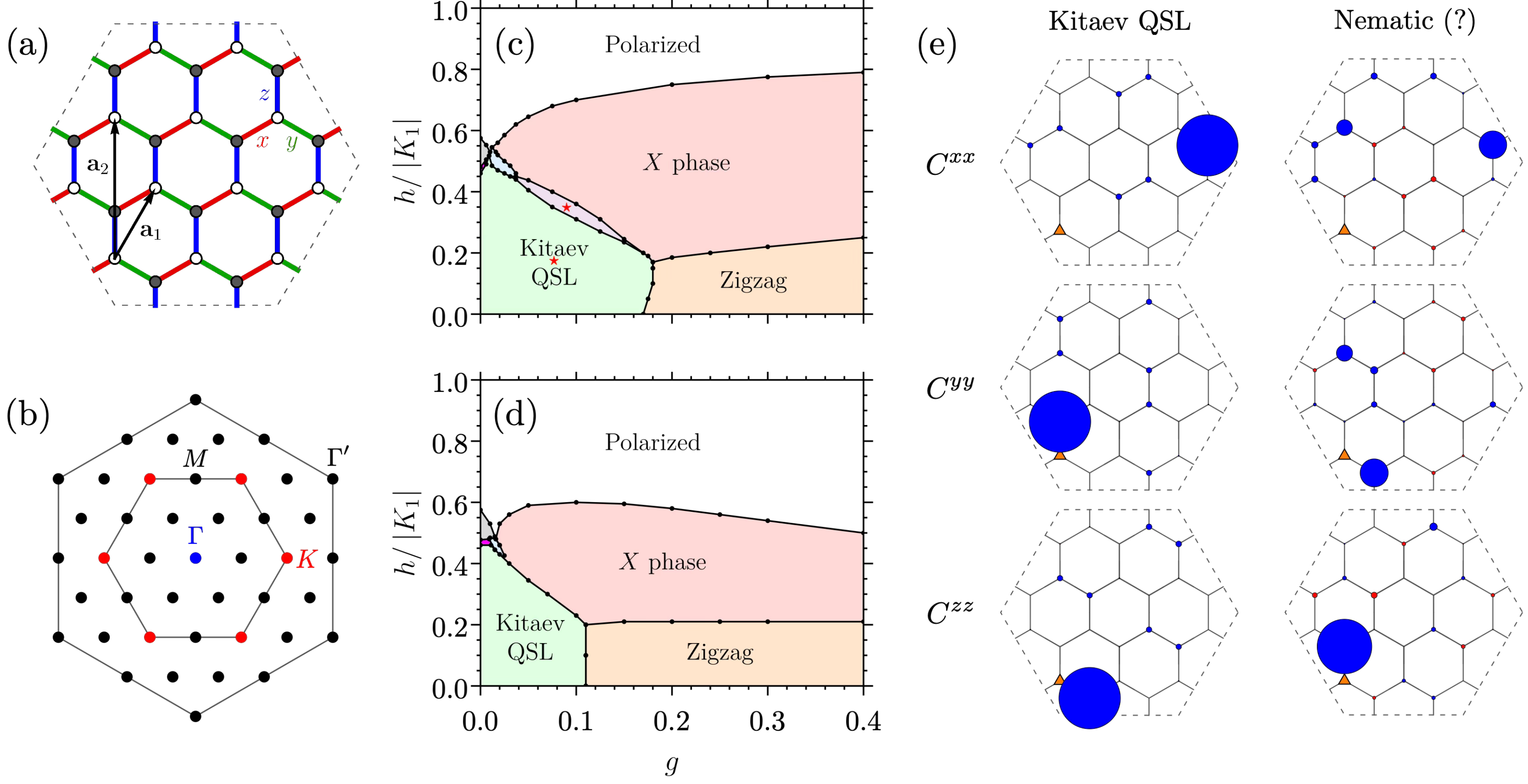} 
    \caption{(a) 24-site cluster used in our ED calculations. 
    (b) Momentum resolution of the finite-size cluster. The inner hexagon represents the first Brillouin zone, whereas the outer hexagon connects the six equivalent $\bm{\Gamma}'$ points.
    (c,d) Ground-state phase diagrams of Models 1 and 2 from Table~\ref{tab:edmodels}. Both models exhibit a robust field-induced $X$ phase which precedes polarizition for nearly all values of $g$ up to $1$.
    (e) Comparison between two-point correlations $C^{\alpha\alpha}$ relative to site $0$ (orange triangle) in the KSL and the purple phase above it in (c). The data were obtained for $(g, h/\abs{K_1}) = (0.075,0.175)$ and $(0.09,0.35)$, which are highlighted in (c). The value of $C^{\alpha\alpha}$ at each site is represented by a circle whose radius is proportional to $\abs{C^{\alpha\alpha}}$ and whose color reflects the sign of $C^{\alpha\alpha}$. Blue (red) circles indicate FM (AFM) correlations.
    }
    \label{fig:ed_cluster+pds}
\end{figure}

To access phenomena away from the vicinity of the pure Kitaev point, we applied Lanczos ED to a cluster with $N=24$ sites, subject to periodic boundary conditions [see Fig.~\ref{fig:ed_cluster+pds}(a)]. Ground-state phase diagrams were then constructed as detailed in the SI. To characterize the physical properties of the phases obtained in our simulations, we computed the static spin structure factor
\begin{equation}
    \mathcal{S} (\mathbf{q}) = \frac{1}{N} \sum_{i,j} e^{\ii \mathbf{q} \cdot (\mathbf{r}_i - \mathbf{r}_j)} \expval{\mathbf{S}_i \cdot \mathbf{S}_j},
    \label{eq:ssf}
\end{equation}
a quantity probed in neutron-scattering experiments, and two-point correlation functions
\begin{equation}
    C^{\alpha\beta} (r) = \expval{S_0^\alpha S_r^\beta}
    \label{eq:correlations}
\end{equation}
with respect to a fixed reference site $0$. In both Eqs.~\eqref{eq:ssf} and \eqref{eq:correlations}, $\expval{\cdots}$ denotes the ground-state expectation value.

The ED phase diagrams of Models 1 and 2 are shown in Figs.~\ref{fig:ed_cluster+pds}(c,d), respectively. At $g=0$, both models are identical up to a rescaling of $K_1$ and undergo a sequence of three phase transitions at fields $h_{c1} \approx 0.4625 \abs{K_1}$, $h_{c2} \approx 0.4775 \abs{K_1}$, and $h_{c3} \approx 0.575 \abs{K_1}$ before entering a (partially) polarized phase with neither symmetry-breaking nor topological order. 
We note that these results are mostly consistent with previous simulations from Ref.~\onlinecite{Hickey_NatComm2019}; the only disagreement lies in the existence of the phase between $h_{c1}$ and $h_{c2}$, which may have been overlooked due to its small extent and different ground-state momentum sector (see SI).
In contrast, when $h=0$, both models exhibit a single phase transition between the KSL and the zigzag phase as $g$ is tuned from $0$ to $1$. The transition occurs at $g_c \approx 0.17$ for Model 1 and at a lower $g_c \approx 0.11$ for Model 2 due to stronger non-Kitaev couplings.% Although one can already infer that the second phase corresponds to zigzag magnetic order by adiabatic continuity to $g=1$, we confirmed that this is the case by analyzing the static structure factor, Eq.~\eqref{eq:ssf}, and two-point correlations, Eq.~\eqref{eq:correlations}, in the ground state.

A comparison between Figs.~\ref{fig:ed_cluster+pds}(c,d) for $g,h \ne 0$ reveals that Model 1 has a higher number of field-induced phases, some of which occur only for $g\ne 0$. Since we cannot exclude that the narrower phases are finite-size effects, we refrain from describing them individually and provide only a brief account of the most interesting case: the purple phase observed in the interval $0.03 \lesssim g \lesssim 0.18$.
This phase is found to have a broad spin structure factor (not shown), and to therefore lack long-range magnetic order. However, as illustrated in Fig.~\ref{fig:ed_cluster+pds}(e), its spin correlations are markedly different from those of a KSL. 
While both phases maximize $\abs{C^{\alpha\alpha}}$ along the corresponding NN $\alpha$-bond\cite{Baskaran_PRL2007}, the purple phase develops stronger correlations along one of the three possible directions ($z$ bonds in the present example).
Hence, in contrast to the KSL, it spontaneously breaks the symmetry of the Hamiltonian under combined threefold rotations of the honeycomb lattice and of the spins around the out-of-plane [111] axis\cite{Janssen_JPCM2019}. Together with the absence of long-range magnetic order, this suggests that the purple phase realizes a form of nematic order.

\begin{figure}[t]
    \center
    \includegraphics[width=0.65\textwidth]{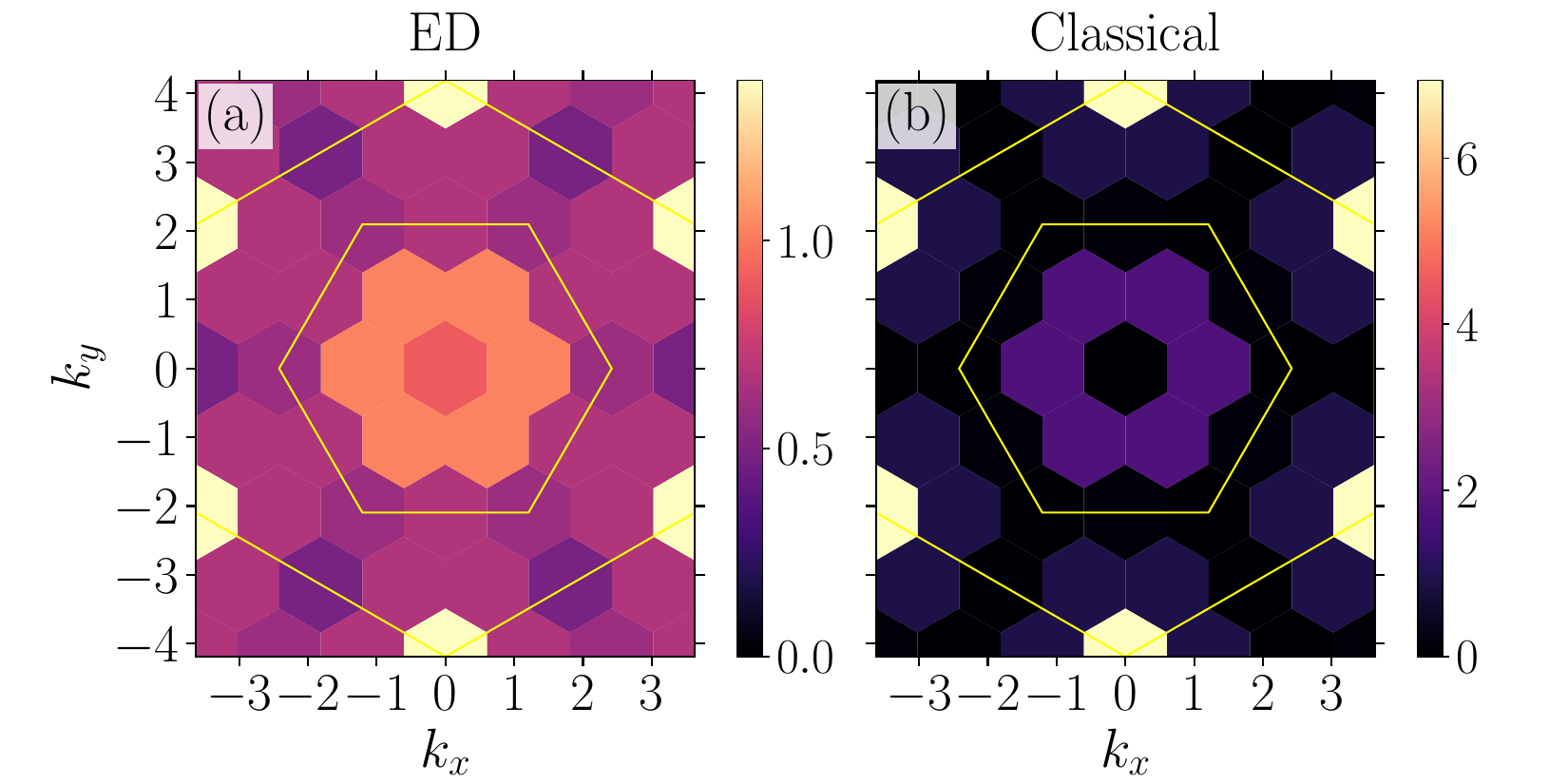} 
    \caption{Static spin structure factor $\mathcal{S} (\mathbf{k})$ obtained by applying (a) ED and (b) a classical energy minimization to Model 1 on the 24-site cluster of Fig.~\ref{fig:ed_cluster+pds}(a). The results were obtained for $(g, h/\abs{K_1}) = (0.75, 0.55)$ and $(0.75,1.18)$, respectively. The yellow hexagons have the same meaning as in Fig.~\ref{fig:ed_cluster+pds}(b).
    }
    \label{fig:ssf_xphase}
\end{figure}

Despite the above differences, the phase diagrams in Figs.~\ref{fig:ed_cluster+pds}(c,d) are dominated by the same \emph{four} phases. Besides the expected Kitaev, zigzag, and field-polarized phases, we encounter a robust, field-induced ``$X$ phase'' that  spans a broad interval $0.02 \lesssim g \le 1$, and therefore extends into the regime that is directly relevant for $\alpha$-RuCl$_3$. Upon increasing $h$, the $X$ phase generally reaches a field-polarized state via a first-order transition.
Its spin structure factor, depicted in Fig.~\ref{fig:ssf_xphase}(a) for a representative point $(g, h/\abs{K_1}) = (0.75,0.55)$, exhibits primary peaks at $\bm{\Gamma}'$ [compare with Fig.~\ref{fig:ed_cluster+pds}(b)] due to the partial polarization along the staggered magnetic field. Additional secondary peaks at the $\bm{K}/2$ points nonetheless suggest the existence of long-range magnetic order. However, with our implementation of ED, we cannot immediately resolve whether this ordered state corresponds to a quantum superposition of degenerate single-$\mathbf{Q}$ spin spirals or a single domain of a more exotic triple-$\mathbf{Q}$ state, where the specification of spin configuration requires three wavevectors $\mathbf{Q}$.

%%%%%%%%%%%%%%%%%%%%%%%%%%%%%%%%%%%%%%%%%%%%%%%%%%%%%%%%%%%%%%%%%%%%%%%%%%%%%%%
% Classical energy minimization
%%%%%%%%%%%%%%%%%%%%%%%%%%%%%%%%%%%%%%%%%%%%%%%%%%%%%%%%%%%%%%%%%%%%%%%%%%%%%%%

To address this issue, we sought a classical analog of the $X$ phase on the same 24-site cluster employed in our ED calculations -- an approach justified by the framework of spin-wave theory\cite{Holstein_PR1940,Igarashi_PRB1992,Mourigal_PRB2010,Consoli_PRB2020}. Using an iterative energy-minimization method (see SI), we searched for ground states of the classical version of Model 1, in which spins $\mathbf{S}_i$ are treated as unit vectors, over a wide range of $g$ and $h$. Among the various phases we encountered, none were single-$\mathbf{Q}$ spirals with an ordering wavevector $\mathbf{Q} = \bm{K}/2$. However, we did obtain a triple-$\mathbf{Q}$ ground state whose structure factor, depicted in Fig.~\ref{fig:ssf_xphase}(b) for $(g, h/\abs{K_1}) = (0.75,1.18)$, shows the same distinctive features as Fig.~\ref{fig:ssf_xphase}(a): primary peaks at $\bm{\Gamma}'$ accompanied by local maxima at $\bm{K}/2$. Altogether, this indicates that the $X$ phase identified within 24-site ED corresponds to a triple-$\mathbf{Q}$ magnetically ordered state. In the SI, we provide further evidence that this is the case by analyzing ED results in a magnetic field that is tilted away from the [111] direction $\unitvec{c}$.

\begin{figure}[t]
    \center
    \includegraphics[width=\textwidth]{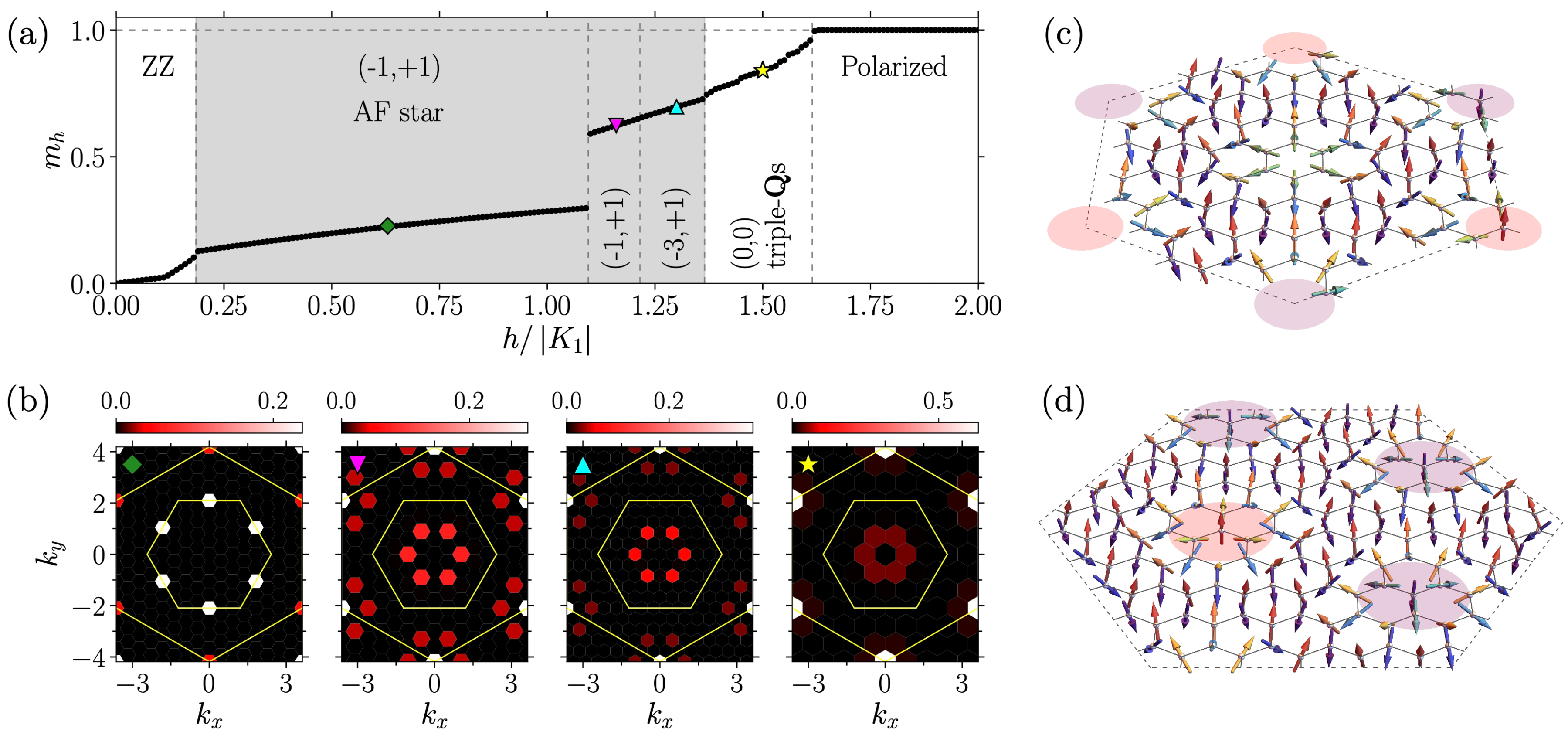} 
    \caption{Results obtained from classical energy minimization for Model 1 at fixed $g = 0.75$.
    (a) Staggered magnetization per site, $m_h = - (Nh)^{-1} \sum_{i} \mathbf{h}_i \cdot \mathbf{S}_i$, as a function of $h$. Between the zigzag (ZZ) and polarized phases, the system realizes a sequence of SkXs (gray shaded region) and topologically trivial triple-$\mathbf{Q}$ phases with progressively smaller ordering wavevectors. The former are characterized by their sublattice-resolved topological charges, $(n_\mathrm{sk}^A, n_\mathrm{sk}^B)$. 
    (b) Static structure factors per site, $\mathcal{S}(\textbf{k})/N$, for the points highlighted on the magnetization curve. Data are shown for the discrete momenta compatible with the clusters on which the solutions were obtained. The inner yellow hexagon delineates the first Brillouin zone.
    (c,d) Magnetic unit cells of the $(-1,+1)$ and $(-3,+1)$ SkXs, respectively. Shaded disks highlight the location of skyrmions.}
    \label{fig:clmin}
\end{figure}

As a next step, we aimed to determine if the above triple-$\mathbf{Q}$ configuration is still a classical ground state of Model 1 for larger system sizes. To this end, we applied the same classical method as before to 22 clusters of varying shapes and sizes. We imposed periodic boundary conditions in all cases and reached a maximum of $N=288$ sites. Working at fixed $g = 0.75$, we then selected the states with the lowest energy per spin for $h/\abs{K_1} \in \left[ 0, 2 \right]$.
The results in Fig.~\ref{fig:clmin} show that the system undergoes a complex magnetization process before entering the field-polarized phase. As in the case of a spatially uniform field, this follows not only from the presence of magnetic frustration, but also from the nontrivial interplay between the magnetic field and the anisotropic spin interactions \cite{Janssen_PRL2016,Janssen_PRB2017,Chern_PRB2017,Janssen_JPCM2019,Chern_PRR2020}. Furthermore, the fact that the classical model has a higher critical field than its spin-$1/2$ counterpart is consistent with previous results showing that quantum fluctuations favor the field-polarized state in a similar system
\cite{Consoli_PRB2020,Consoli_JPCS2022}.

Upon increasing $h$, the low-field zigzag (ZZ) phase is suppressed at $h/\abs{K_1} \approx 0.19$, giving way to a sequence of three AFM skyrmion crystals (SkXs). These are triple-$\mathbf{Q}$ spin textures that cannot be smoothly deformed into a collinear Néel state and are characterized by a pair of integer-valued topological invariants, $n_\mathrm{sk}^A$ and $n_\mathrm{sk}^B$, which reflect the number of times the spin configurations on the $A$ and $B$ sublattices wrap the unit sphere\cite{Nagaosa2013} (see SI for details).

The first SkX occurs within the extended interval $0.19 < h/\abs{K_1} < 1.09$ and has opposite topological charges on each sublattice, $(n_\mathrm{sk}^A, n_\mathrm{sk}^B) = (-1,+1)$. Its structure factor exhibits pronounced Bragg peaks at the three inequivalent $\bm{M}$ points [see Fig.~\ref{fig:clmin}(b), left]. By inspecting the spin configuration directly in real space, we find that it has an 8-site magnetic unit cell and corresponds to a state dubbed AFM star in an early study of the classical Heisenberg-Kitaev model in a \emph{uniform} magnetic field \cite{Janssen_PRL2016}.
Later on, this same type of order was suggested as a potential ground state of the Kitaev material Na$_2$Co$_2$TeO$_6$ in zero field\cite{Chen_PRB2021,Krueger_PRL2023,Gu_CPL2025,Francini_PRB2024,Francini_PRB2025}.
The absence of this AFM star phase from the quantum phase diagrams of Figs.~\ref{fig:ed_cluster+pds}(c,d) is intriguing, given that it occupies a broad region in Fig.~\ref{fig:clmin} and is commensurate with the 24-site cluster in Fig.~\ref{fig:ed_cluster+pds}(a). This may be due to the fragility of the state to quantum fluctuations, which was previously demonstrated within spin-wave theory\cite{Consoli_PRB2020}.

The second and third SkXs, henceforth labeled SkX$_2$ and SkX$_3$, are built from a linear combination of ordering wavevectors $\mathbf{Q} = 3\bm{K}/7 \approx 0.43\bm{K}$ and $\mathbf{Q} = 0.4\bm{K}$, respectively. Thus, they have larger magnetic unit cells than an AFM star state [see Figs.~\ref{fig:clmin}(c,d)] and, given the greater proximity to $\mathbf{Q} = \bm{K}/2$, are better candidates for a classical analog of the $X$ phase.
Similarly to the AFM star, SkX$_2$ is an AFM SkX with topological charges $(n_\mathrm{sk}^A, n_\mathrm{sk}^B) = (-1,+1)$. SkX$_3$, on the other hand, is characterized by two different, yet equally favorable, sets of integers: $(n_\mathrm{sk}^A, n_\mathrm{sk}^B) = (-3,+1)$ and $(-1,+3)$. This indicates that SkX$_3$ \emph{spontaneously} develops a net topological charge of $\pm 2$ per magnetic unit cell, a behavior reminiscent of the ``ferrichiral'' skyrmion phase identified in a recent study of an extended Kitaev model on the triangular lattice\cite{Wen_2026}.

Finally, the system realizes a sequence of topologically trivial ($n_\mathrm{sk}^A = n_\mathrm{sk}^B = 0$) triple-$\mathbf{Q}$ states between $h/\abs{K_1} = 1.37$ and $1.62$ before reaching the field-polarized phase. We refrain from describing these states in detail, but note that their spatial period increases consistently with $h$. This extends a trend already seen among the three aforementioned SkXs and hints at the possible existence of an even greater number of triple-$\mathbf{Q}$ phases, both topological and trivial, in the thermodynamic limit.
It also implies that of all classical phases obtained here, the one whose ordering wavevectors are closest to $\bm{K}/2$ is SkX$_2$. Given the limited momentum resolution of our 24-site ED, this does not unambiguously identify the $X$ phase as SkX$_2$ -- in fact, the $X$ phase may well correspond to multiple phases in the thermodynamic limit. Our results nevertheless signal that applying staggered magnetic fields to Kitaev materials may be a promising route to realize SkXs with periodicities smaller than those of their metallic counterparts stabilized by the Dzyaloshinskii–Moriya interaction\cite{Fert_NatRevMat2017}.

Given that the SkX$_2$ and SkX$_3$ states would have periodicities on the order of $\sim10$~nm in $\alpha$-RuCl$_3$, they may be detectable by magnetic exchange force microscopy\cite{Kaiser_Nature2007,Grenz_PRL2017}, an imaging technique with a high spatial resolution of $\lesssim\!1$~nm that is applicable to thin insulating films. An alternative method would be Lorentz transmission electron microscopy\cite{Seki_Science2012}, although its $\sim\!10$~nm resolution is likely insufficient for the systems at hand. 
The SkX$_2$ and SkX$_3$ phases may also display topological magnon bands and an associated thermal Hall effect\cite{Akazawa_PRR2022,Takeda_NatComm2024,Kawano_2025}, which is nonetheless non-quantized and therefore does not necessarily provide a smoking-gun signature of SkXs.

%%%%%%%%%%%%%%%%%%%%%%%%%%%%%%%%%%%%%%%%%%%%%%%%%%%%%%%%%%%%%%%%%%%%%%%%%%%%%%%
% First principles 
%%%%%%%%%%%%%%%%%%%%%%%%%%%%%%%%%%%%%%%%%%%%%%%%%%%%%%%%%%%%%%%%%%%%%%%%%%%%%%%

\begin{figure}[t]
    \centering
    \includegraphics[width=\linewidth]{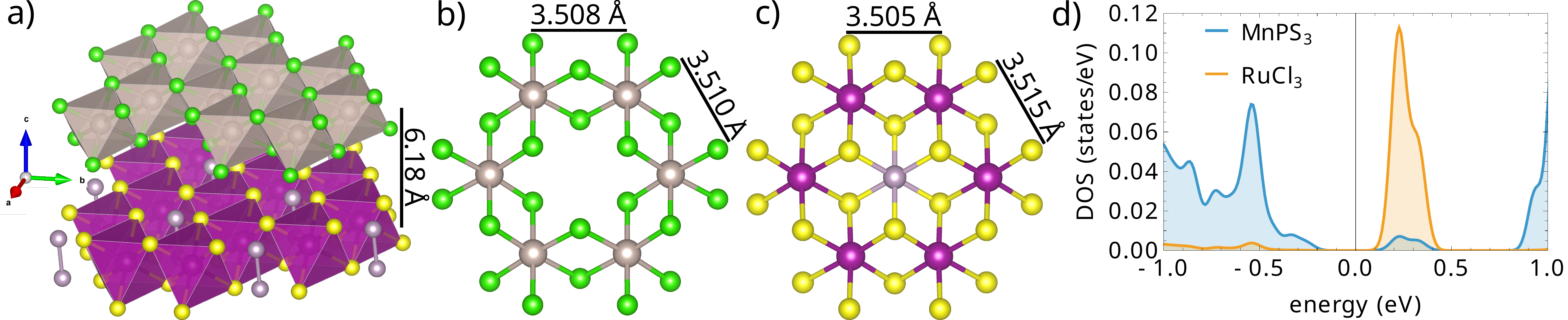}
    \caption{(a) Structure of the bilayer $\alpha$-RuCl$_3$/MnPS$_3$ showing the interlayer cation-cation distance. Ru:gray, Cl:green, Mn:purple, S:yellow, P:silver (b,c) Single layers of $\alpha$-RuCl$_3$ and MnPS$_3$, respectively, showing cation-cation distances. (d) Density of states for a fully FM configuration (interlayer AFM configuration is qualitatively similar).}
    \label{fig:structure}
\end{figure}

As mentioned above, a plausible materials platform to scrutinize a Kitaev material subject to a \emph{staggered} magnetic field is represented by a bilayer of $\alpha$-RuCl$_3$ and AFM MnPS$_3$. 
To obtain a characteristic coupling scale for this heterobilayer, we use density functional theory calculations\cite{VASP,VASP-PAW} including spin-orbit coupling, an on-site Coulomb repulsion $U$ applied to the Mn and Ru-$d$ electrons \cite{Dudarev1998}, and the DFT-D3 van der Waals correction\cite{DFTD3}. 
We start by building the structure of a heterobilayer as shown in Fig.~\ref{fig:structure}(a) that is subsequently optimized. This AA stacking structure, with the Mn atoms directly above the Ru atoms, is necessary to realize our effective field in the Hamiltonian \eqref{eq:Hparts}. Any alternate stacking would modify the sign structure or modulate the magnitude of the effective field, or for Mn atoms placed equidistant from Ru ones, potentially average to zero effective field.
Further details about the methodology, including the structural relaxation, are provided in the SI. Figures~\ref{fig:structure}(b,c) show relevant bond lengths for $\alpha$-RuCl$_3$ and MnPS$_3$ in our relaxed structure, reflecting that they are closely lattice-matched. We subsequently use the optimized structure to calculate the energy difference between two magnetic states both with FM order within the layers and differing by the alignment or anti-alignment between the layers. Importantly, there is no charge transfer between the layers and the system remains in an insulating state regardless of the interlayer orientation, as shown in Fig.~\ref{fig:structure}(d). We obtain an effective field $h \simeq 0.3$ meV, which alone is not enough to enable access to the $X$ phase. However, applied pressure, causing a compression of the layers along the $c$ axis, has widely been used in van der Waals materials to tune interlayer couplings \cite{Jiao_2023,Xia2021,Kapeghian2024,Bag2024}. Based on calculations where external pressure is applied, as detailed in the SI, it should be possible to access this phase at $\sim$ 2 GPa (corresponding to a 0.1~\AA ~reduction in interlayer distance), which gives rise to an effective field of $h\simeq 3.2$ meV. This translates to a coupling of $h/|K_1| \simeq 0.6,0.4$ in models 1 and 2, respectively, within the parameter region of the $X$ phase for most values of $g$. Overall, our calculations argue for the experimental feasibility of using pressure to access the novel phase we have found.

%%%%%%%%%%%%%%%%%%%%%%%%%%%%%%%%%%%%%%%%%%%%%%%%%%%%%%%%%%%%%%%%%%%%%%%%%%%%%%%
% Conclusion
%%%%%%%%%%%%%%%%%%%%%%%%%%%%%%%%%%%%%%%%%%%%%%%%%%%%%%%%%%%%%%%%%%%%%%%%%%%%%%%

In summary, we studied two-dimensional extended Kitaev models subject to an effective staggered magnetic field. We showed that, in addition to inducing Majorana Fermi surfaces and \emph{antichiral} edge modes in the KSL, the staggered field can give rise to a nonmagnetic nematic phase and to SkXs, both with and without a net topological charge.
To complement these results, we employed first-principle simulations to assess the prospect of realizing these phenomena in an $\alpha$-RuCl$_3$/MnPS$_3$ bilayer. Our DFT calculations demonstrate that this heterostructure is approximately lattice-matched and MnPS$_3$ leads to an effective field of about $h \simeq 0.3$ meV.

A compelling direction for future research is to consider how a Kitaev material is influenced by a proximity-induced field $\mathbf{h}_i$ with the periodicity of a moiré superlattice. This would account for the lattice mismatch in the proposed $\alpha$-RuCl$_3$/MnPS$_3$ heterostructure, which we estimate to be of up to $1.5\%$ based on the \emph{bulk} lattice constants of these materials\cite{Cao_PRB2016,OUVRARD1985}. Since the resulting moiré unit cell has a linear dimension of roughly 70 lattice spacings ($\sim25$~nm), the SkX states discussed here should represent good candidates for the local ordering in the AA-stacked regions of the heterostructure. However, the configuration will certainly change near regions with AB stacking. This opens the door to realize different forms of magnetic order\cite{Kasra_PNAS2020,Akram_NanoLett2024}, especially in the vicinity of the field-induced phase boundaries of our ED phase diagrams. The moiré periodicity should likewise have a nontrivial effect on the antichiral KSL.
On the materials front, it would be worthwhile not only to pursue an experimental realization of the $\alpha$-RuCl$_3$/MnPS$_3$ bilayer, but also to seek other material platforms capable of reproducing the physics discussed here.

%%%%%%%%%%%%%%%%%%%%%%%%%%%%%%%%%%%%%%%%%%%%%%%%%%%%%%%%%%%%%%%%%%%%%%%%%%%%%%%
%%%%%%%%%%%%%%%%%%%%%%%%%%%%%%%%%%%%%%%%%%%%%%%%%%%%%%%%%%%%%%%%%%%%%%%%%%%%%%%

\section{Associated Content}

\noindent {\bf Supporting Information}

\noindent Contains further details on the perturbative calculation around the Kitaev limit, exact diagonalization, iterative energy-minimization algorithm, and computation of the topological charges and details on the density-functional theory calculations. Includes Refs.~\cite{Lieb_PRL1994,Panigrahi_PRB2023,Walker_PRL1977,Berg_NPB1981,Diaz_PRL2019,Nagaosa2013,VASP,VASP-PAW,PBE,DFTD3,Kee_2016,DFT-ulg,Steinmann2011_1,Steinmann2011_2,Tkatchenko2009}.

%%%%%%%%%%%%%%%%%%%%%%%%%%%%%%%%%%%%%%%%%%%%%%%%%%%%%%%%%%%%%%%%%%%%%%%%%%%%%%%
%%%%%%%%%%%%%%%%%%%%%%%%%%%%%%%%%%%%%%%%%%%%%%%%%%%%%%%%%%%%%%%%%%%%%%%%%%%%%%%

\section{Author Information}
\noindent \textbf{Corresponding Author}: Pedro M. Cônsoli (pconsoli@asu.edu), Onur Erten\\ (onur.erten@asu.edu) \& Antia Botana (asanc157@asu.edu)

\noindent {\bf Author Contributions:} PMC, JK, ASB, and OE conceived and designed the project. PMC performed the exact-diagonalization and classical simulations. EDR and ASB carried out the first-principles calculations. All authors analyzed the results and contributed to writing the manuscript.

\noindent {\bf Notes:} The authors declare no competing financial interest.

%%%%%%%%%%%%%%%%%%%%%%%%%%%%%%%%%%%%%%%%%%%%%%%%%%%%%%%%%%%%%%%%%%%%%%%%%%%%%%%

\section{Acknowledgements}

We thank Hui-Ke Jin for preliminary DMRG calculations, as well as Eric C. Andrade and Natalia Perkins for helpful discussions. P.M.C. is also grateful to Eric C. Andrade, Niccol\`o Francini, Lukas Janssen, and Matthias Vojta for collaborations on previous related projects. 
This work is supported by the U.S. Department of Energy, Office of Science, Office of Basic Energy Sciences, Material Sciences and Engineering Division under Award Number DE-SC0025247. EDR and ASB acknowledge the support of NSF Grant No. DMR-2206987.
J. K. acknowledges support from the Deutsche Forschungsgemeinschaft (DFG, German Research Foundation) grant TRR 360 - 492547816 [14]. J.K. further acknowledges support from DFG under Germany’s Excellence Strategy (EXC-2111-390814868), DFG Grants No. KN1254/1-2, KN1254/2-1 and SFB 1143 (project-id 247310070), as well as the Munich Quantum Valley, which is supported by the Bavarian state government with funds from the Hightech Agenda Bayern Plus. J.K. further acknowledges support from the Imperial-TUM flagship partnership, as well as the Keck Foundation.
The authors acknowledge Research Computing at Arizona State University for providing HPC resources\cite{Sol} that contributed to the results reported in this paper.

 %%%%%%%%%%%%%%%%%%%%%%%%%%%%%%%%%%%%%%%%%%%%%%%%%%%%%%%%%%%%%%%%%%%%%%%%%%%%%%%

\bibliography{references}
\end{document}

\setstretch{1.2}

In this supplementary file, we provide additional details on four aspects discussed in the main text. First, we outline the perturbative calculation we applied to study the effect of a weak staggered field on a Kitaev spin liquid (KSL). Second, we provide additional information on our ED calculations, including data to exemplify how we extracted the phase boundaries in Figs.~2(c) and 2(d) of the main text. Third, we describe the iterative energy-minimization algorithm we applied to obtain classical ground states of Model 1. Fourth, we explain how we computed the topological charges $n_\mathrm{sk}^\mu$ used to characterize the skyrmion crystals uncovered in our classical simulations. And finally, we provide details of the first principles calculations.

%%%%%%%%%%%%%%%%%%%%%%%%%%%%%%%%%%%%%%%%%%%%%%%%%%%%%%%%%%%%%%%%%%%%%%%%%%%%%%%
\section{\normalsize Perturbation theory in $h$ around the Kitaev limit}
%%%%%%%%%%%%%%%%%%%%%%%%%%%%%%%%%%%%%%%%%%%%%%%%%%%%%%%%%%%%%%%%%%%%%%%%%%%%%%%

In this section, we show how perturbation theory can be used to infer that a KSL subject to a weak staggered magnetic field develops Majorana Fermi surfaces and antichiral edge modes. This calculation is closely related to the one presented in Kitaev's original paper\cite{kitaev2006anyons} and to Ref.~\onlinecite{Nakazawa_PRB2022}.

We start by separating the full Hamiltonian $\mathcal{H} = \mathcal{H}_0 + \mathcal{V}$ into an unperturbed Kitaev Hamiltonian, $\mathcal{H}_0$, and the Zeeman term, $\mathcal{V}$:
\begin{align}
    \HH_0 &= \sum_{\gamma} K_\gamma \sum_{\langle ij \rangle_\gamma } \sigma_i^\gamma \sigma_j^\gamma,
    \notag \\
    \mathcal{V} &= -\sum_{i} \vec{h}_i \cdot \bm{\sigma}_i
    = - \sum_{i} (-1)^i \sum_{\gamma} h_\gamma \sigma_i^\gamma.
\end{align}
As usual, the model is defined on a honeycomb lattice with nearest-neighbor bonds of types $\gamma = x,y,z$ and sublattices $A$ and $B$. Note that, differently from the main text, we now allow for an arbitrary staggered field $\vec{h}_i = (-1)^i \vec{h}$. Furthermore, we will work directly with Pauli matrices $\sigma_i^\gamma$ instead of spin-$1/2$ operators $S_i^\gamma = \hbar \sigma_i^\gamma/2$ for convenience.
Unless stated otherwise, we will assume that the Kitaev couplings are isotropic ($K_\gamma = K$) for simplicity. However, our results are valid for the entire ``$B$ phase'' of the Kitaev model\cite{kitaev2006anyons}, which is realized when $\abs{K_\alpha} \le \abs{K_\beta} + \abs{K_\gamma}$ for all permutations $(\alpha,\beta,\gamma)$ of $(x,y,z)$.

Before applying perturbation theory, let us briefly review some of the main ingredients behind Kitaev's exact solution of $\HH_0$. Following Ref.~\onlinecite{kitaev2006anyons}, we can represent each spin-$1/2$ degree of freedom $\bm{\sigma}_i$ in terms of four Majorana fermions $\left\{ b_i^x, b_i^y, b_i^z, c_i \right\}$ obeying the anticommutation relations
$\acomm{b_i^\alpha}{b_j^\beta} = \delta_{ij} \delta_{\alpha\beta}$,
$\acomm{b_i^\alpha}{c_j} = 0$,
and $\acomm{c_i}{c_j} = \delta_{ij}$.
With this, the representation $\sigma_i^\gamma = \ii b_i^\gamma c_i$ reproduces the appropriate SU(2) spin algebra as long as one imposes the constraint $D_i = b_i^x b_i^y b_i^z c_i = 1$. The unperturbed Hamiltonian then reads
\begin{equation}
    \HH_0 = -\ii \sum_\gamma K_\gamma \sum_{\langle ij \rangle_\gamma} \hat{u}_{ij}^\gamma c_i c_j,
\end{equation}
where $\hat{u}_{ij}^\gamma = \ii b_{i}^\gamma b_{j}^\gamma = - \hat{u}_{ji}^\gamma$ is an operator defined for sites $i$ and $j$ that belong to the $A$ and $B$ sublattices, respectively, and are connected by a $\gamma$-bond. Since these operators commute with each other and with $\HH_0$, one can divide the unperturbed Hamiltonian into separate gauge sectors, in which the $\hat{u}_{ij}^\gamma$ operators are replaced by a set of integer eigenvalues $u_{ij}^\gamma = \pm1$. 

However, as in any gauge theory, this description is redundant. Two gauge sectors are only inequivalent if they realize different sets of (gauge-invariant) flux eigenvalues $\left\{ w_p \right\}$, where $w_p = \prod_{(ij) \in p} u_{ij}^\gamma = \pm1$ is given by the product of the $u_{ij}^\gamma$ variables around a given elementary hexagonal plaquette $p$. It then follows from Lieb's theorem\cite{Lieb_PRL1994} that the ground state of the system can be obtained by diagonalizing $\HH_0$ within any gauge sector $\left\{ u_{ij}^\gamma \right\}$ that yields $w_p = +1$ for every $p$.
By doing so, one also finds that the excitations of a KSL with isotropic couplings (or, more generally, of the $B$ phase alluded to above) are gapless Majorana fermions and gapped visons. The latter correspond to fluxes $w_p = -1$, which can only be created in pairs at an energy cost of $\Delta_{2f} \sim \abs{K}$.

Since visons are gapped excitations, the low-energy physics of the full Hamiltonian $\HH$ is constrained to the zero-flux subspace $\mathcal{L}_0$ as long as $h \ll \Delta_{2f}$. If $P$ is a projector onto $\mathcal{L}_0$, then the effective low-energy Hamiltonian can be derived via a standard perturbative expansion:
\begin{align}
    \mathcal{H}_\mathrm{eff} 
    &\equiv \sum_{n=0}^{\infty} \mathcal{H}_\mathrm{eff}^{(n)}
    %\notag \\
    = \mathcal{H}_0 + P \mathcal{V} \left( \sum_{n=0}^{\infty} R^{n} \right) P,
    \label{eq:Heff expansion}
\end{align}
where $R = (1-P)(E-\mathcal{H}_0)^{-1} (1-P) \mathcal{V}$. Following Ref.~\onlinecite{kitaev2006anyons}, we will use the approximation $R \approx -\Delta_{2f}^{-1} (1-P) \mathcal{V}$, which yields the correct qualitative results for the lowest orders of the perturbation series.

The first-order term $\mathcal{H}_\mathrm{eff}^{(1)} = P\mathcal{V} P$ vanishes identically because $\mathcal{V}$ only connects states with different flux configurations. This result can be verified by noting that a single Pauli matrix $\sigma^\alpha_{\gamma}$ commutes with all but one of the $\hat{u}_{ij}$ operators. The exception is the operator $\hat{u}_{mn}^\gamma$, for which $\acomm{\sigma^\alpha_{m}}{\hat{u}_{mn}^\gamma} = 0$. 
It then follows that if $\ket{ \psi_{\left\{ u_{ij} \right\}} }$ is a state belonging to a gauge sector $\left\{ u_{ij} \right\}$, $\sigma^\alpha_{m}\ket{ \psi_{\left\{ u_{ij} \right\}} }$ will be in a gauge sector $\left\{ \tilde{u}_{ij} \right\}$ which only differs from $\left\{ u_{ij} \right\}$ in that $\tilde{u}_{mn} = -u_{mn}$.
Consequently, $\ket{ \psi_{\left\{ u_{ij} \right\}} }$ and $\sigma^\alpha_{m}\ket{ \psi_{\left\{ u_{ij} \right\}} }$ have opposite fluxes on the two hexagonal plaquettes joined by the $(mn)$ bond.

The second-order term $\mathcal{H}_\mathrm{eff}^{(2)} \approx -\Delta_{2f}^{-1} P \mathcal{V}^2 P$ is the first nonvanishing contribution to the perturbation series. One can verify that, apart from shifting the Hamiltonian by a constant, its only effect is to renormalize the Kitaev couplings: $K_\gamma \longmapsto \tilde{K}_\gamma = K + ( h_\gamma^2/\Delta_{2f} )$. Since the correction $(h_\gamma^2/\Delta_{2f})$ is small compared to $K$, this renormalization is innocuous: the system remains in the aforementioned $B$ phase, without suffering any qualitative changes.

\begin{figure}[t]
    \center
    \includegraphics[width=0.5\textwidth]{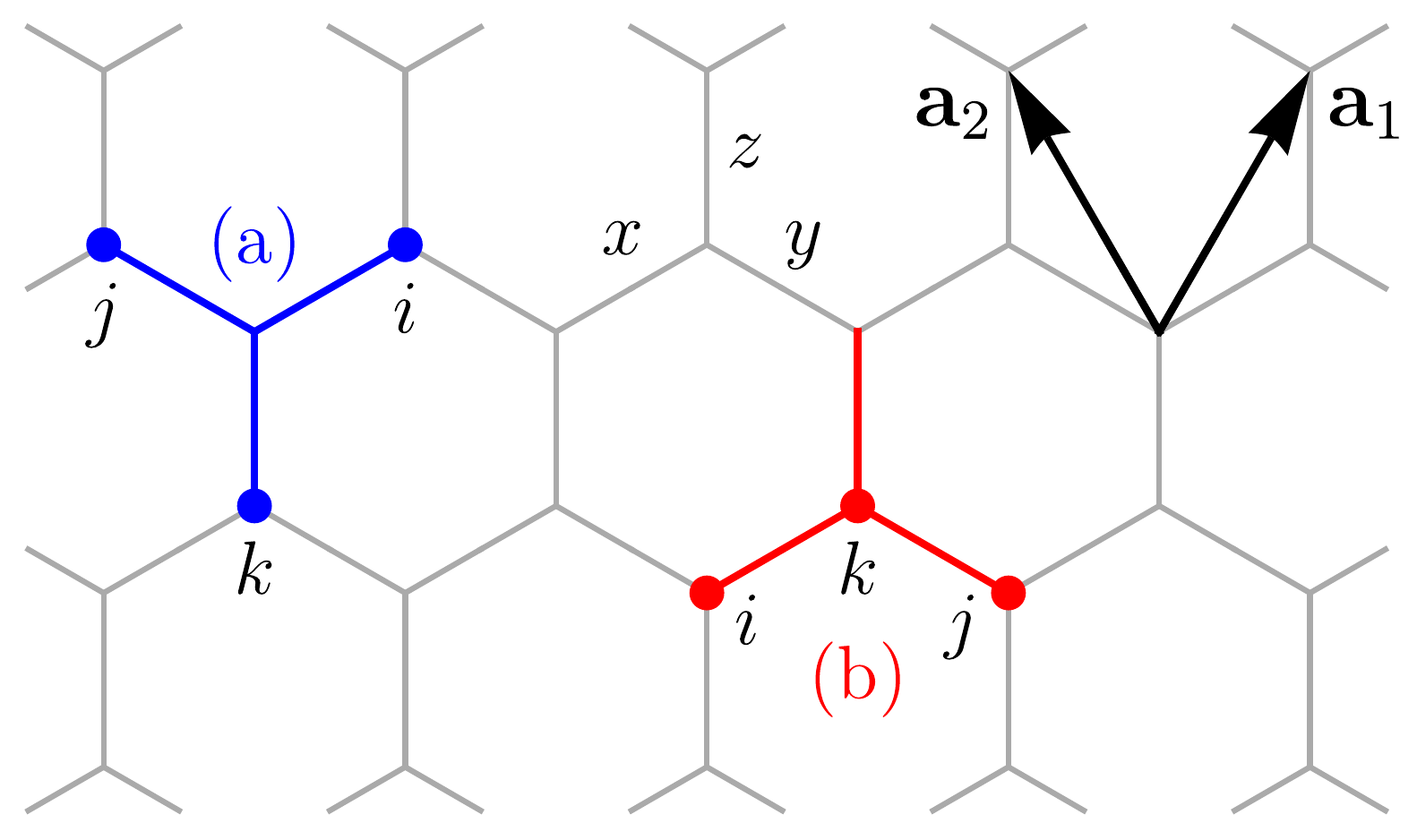} 
    \caption{Schematic depicting the sites, $i$, $j$, and $k$, involved in the two types of processes, (a) and (b), which yield a nonzero contribution to the third-order term of the perturbation series. The blue and red lines highlight the bonds whose gauge variables $u_{ij}^\gamma$ are flipped after the action of the three-body term $\sigma_i^x \sigma_j^y \sigma_k^z$.
    Also shown are the primitive vectors $\vec{a}_1 = (\sqrt{3},3)/2$ and $\vec{a}_2 = (-\sqrt{3},3)/2$ that enter Eqs.~\eqref{eq:fq} and \eqref{eq:Deltaq}.}
    \label{fig:ptsupp}
\end{figure}

The first significant change in the low-energy physics of the system comes from
\begin{equation}
    \HH_\mathrm{eff}^{(3)} \approx -\frac{1}{\Delta_{2f}^2} 
    \sum_{i\alpha} \sum_{j\beta} \sum_{k\gamma} (-1)^{i+j+k} h_\alpha h_\beta h_{\gamma} P \sigma_i^\alpha \sigma_j^\beta \sigma_k^\gamma P.
    \label{eq:Heff3}
\end{equation}
As in Kitaev's original paper\cite{kitaev2006anyons}, the projected three-body interaction $P \sigma_i^\alpha \sigma_j^\beta \sigma_k^\gamma P$ can only be nonzero if $\alpha \ne \beta \ne \gamma$. When $(\alpha, \beta, \gamma) = (x,y,z)$, the nonvanishing contributions are due to the spatial configurations (a) and (b) in Fig.~\ref{fig:ptsupp}, both of which conserve all $w_p$ eigenvalues because they flip an \emph{even} number of $u_{ij}^\gamma$ variables per plaquette. Similar symmetry-related arrangements of $(i,j,k)$ follow for the other combinations of $(\alpha,\beta,\gamma)$.

When expressed in terms of Majorana fermions, third-order contributions of type (a) are found to be \emph{quartic} in the itinerant $c_i$ Majoranas. Thus, they do not immediately change the Majorana spectrum and can be neglected. In contrast, type-(b) terms yield $\sigma_i^x \sigma_j^y \sigma_k^z = -\ii \hat{u}_{ki}^x \hat{u}_{kj}^y c_i c_j D_k$, and therefore map onto a next-nearest-neighbor hopping mediated by the gauge fields.

Hence, by collecting all terms up to third order in Eq.~\eqref{eq:Heff expansion} and choosing a zero-flux gauge with $u_{ij}^\gamma = +1$ on every bond, we obtain the effective Hamiltonian
\begin{equation}
    \HH_\mathrm{eff} = \ii \left( 
    \sum_\gamma \tilde{K}_\gamma \sum_{\langle ij \rangle_\gamma} c_i c_j
    +
    \kappa \sum_{\llangle ij \rrangle} v_{ij} c_i c_j
    \right),
    \label{eq:Heff}
\end{equation}
where $\kappa = (h_x h_y h_z)/\Delta_{2f}^2$ and $v_{ij} = \pm 1$ have the sign structure depicted in Fig. 1(b) of the main text. After a Fourier transform, Eq.~\eqref{eq:Heff} becomes
\begin{equation}
    \HH_\mathrm{eff} = \sum_{\vec{q}}^{\mathrm{HBZ}}
    \begin{pmatrix}
        c_{\vec{q}A}^\dagger & c_{\vec{q}B}^\dagger
    \end{pmatrix}
    \begin{pmatrix}
        \Delta(\vec{q}) & \ii f(\vec{q})
        \\
        -\ii f^*(\vec{q}) & \Delta(\vec{q})
    \end{pmatrix}
    \begin{pmatrix}
        c_{\vec{q}A} \\
        c_{\vec{q}B}
    \end{pmatrix},
    \label{eq:Heff q}
\end{equation}
with the sum restricted to half of the Brillouin zone (HBZ). The functions
\begin{align}
    f(\vec{q}) &= \tilde{K}_x e^{\ii \vec{q} \cdot \vec{a}_1} + \tilde{K}_y e^{\ii \vec{q} \cdot \vec{a}_2} + \tilde{K}_z,
    \label{eq:fq}
    \\
    \Delta(\vec{q}) &= 2\kappa \left\{ 
    \sin (\vec{q} \cdot \vec{a}_1) + \sin (\vec{q} \cdot \vec{a}_2) + \sin\left[ \vec{q} \cdot (\vec{a}_1 - \vec{a}_2) \right]
    \right\}.
    \label{eq:Deltaq}
\end{align}
are defined in terms of the primitive vectors of the honeycomb lattice, $\vec{a}_{1,2} = (\pm\sqrt{3}/2, 3/2)$, shown in Fig.~\ref{fig:ptsupp}.
The spectrum of $\HH_\mathrm{eff}$ follows from the diagonalization of the $2\times 2$ matrix in Eq.~\eqref{eq:Heff q}:
\begin{equation}
    \epsilon_{\pm} (\vec{q}) = \Delta(\vec{q}) \pm \abs{ f(\vec{q}) }.
\end{equation}
Since the renormalized Kitaev couplings $\tilde{K}_\gamma = K + h_\gamma^2/\Delta_{2f}$ have but a small anisotropy, the function $\abs{f(\vec{q})}$ still has zeros at momenta $\pm \vec{q}^*$ in the first Brillouin zone\cite{kitaev2006anyons}. Were it not for $\Delta(\vec{q})$, this would give rise to a pair of Dirac cones at zero energy. However, since $\Delta(-\vec{q}) = -\Delta(\vec{q})$, the Dirac cones are now subject to opposite energy shifts $\pm \abs{\Delta(\vec{q}^*)}$, as illustrated in Fig.~1(b) in the main text.
When $\vec{h} \parallel [111]$ as in the main text, we find that $\vec{q}^* = K = 4\pi/(3\sqrt{3}) \unitvec{x}$ and the splitting $\abs{\Delta(\vec{q^*})} = 3\sqrt{3} \kappa$.

%%%%%%%%%%%%%%%%%%%%%%%%%%%%%%%%%%%%%%%%%%%%%%%%%%%%%%%%%%%%%%%%%%%%%%%%%%%%%%%%%%%%%%
\section{\normalsize Antichiral edge modes on a ribbon with zigzag edges}
%%%%%%%%%%%%%%%%%%%%%%%%%%%%%%%%%%%%%%%%%%%%%%%%%%%%%%%%%%%%%%%%%%%%%%%%%%%%%%%%%%%%%%

\begin{figure}[t]
    \center
    \includegraphics[width=0.65\textwidth]{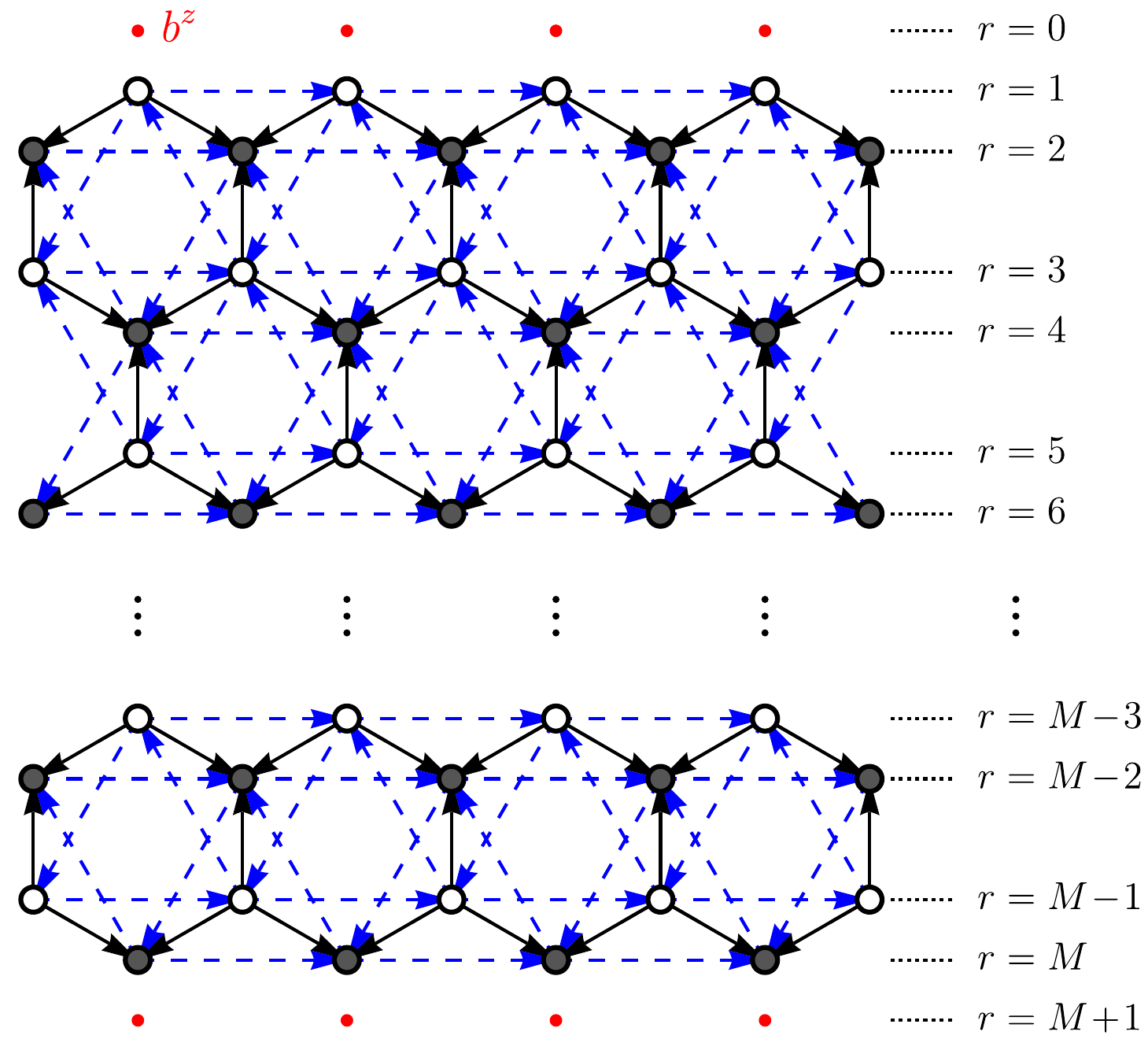} 
    \caption{Illustration of the ribbon geometry utilized in the calculation that resulted in Figs. 1(c) and 1(d) of the main text. We impose periodic boundary conditions along the $x$ (horizontal) direction, while keeping open boundary conditions along $y$. Full and dashed arrows indicate nearest-neighbor and next-nearest-neighbor hopping amplitudes, respectively. The red dots on rows $r=0$ and $r=M+1$ represent dangling Majorana fermions that must be taken into account in the computation of the spectrum.}
    \label{fig:zzribbon}
\end{figure}

To substantiate the claim that, like the modified Haldane model of Ref.~\onlinecite{Colomes_PRL2018}, a Kitaev spin liquid develops antichiral edge modes in the presence of a weak staggered magnetic field, we examine an augmented Kitaev model on a ribbon with zigzag edges. Given the geometry of this system, it is convenient to label the positions $(r,x)$ of the degrees of freedom in terms of a row index $r = 0,1,\ldots, M+1$ and the $x$ coordinate. As illustrated in Fig.~\ref{fig:zzribbon}, rows with odd (even) $0<r\le M$ intersect sites on the $A$ ($B$) sublattice of the honeycomb lattice, whereas the outermost rows, $r=0$ and $r=M+1$, host dangling Majorana fermions. For concreteness, we assume that $M$ is even, i.e., we have an equal number of $A$ and $B$ rows. Furthermore, we impose periodic boundary conditions along the $x$ direction, specialize our discussion to the case with isotropic Kitaev couplings, $K_\gamma = K$, and work in units where the primitive vector $\vec{a}_x$ in Fig.~\ref{fig:zzribbon} has unit length and $K=1$.

The full effective Hamiltonian can then be written as $\HH = \HH_\mathrm{bulk} + \HH_\mathrm{edges}$ with
\begin{align}
    \HH_\mathrm{bulk} &= \ii K \sum_x \left[ \sum_{r=1}^{M/2} c_{2r-1,x} \left( c_{2r,x-1/2} + c_{2r,x+1/2} \right) + \sum_{r=3}^{M/2} c_{2r-1,x} c_{2r-2,x} \right]
    \notag \\
    &+ \ii\kappa \sum_x \left[ \sum_{r=1}^{M-2} c_{rx} \left( c_{r,x+1} + c_{r+2,x-1/2} - c_{r+2,x+1/2} \right) + \sum_{r=M-1}^M c_{rx} c_{r,x+1} \right]
\end{align}
and
\begin{align}
    \HH_\mathrm{edges} &= -h_z \sum_x \left( \sigma_{1x}^z - \sigma_{Mx}^z \right)
    = -\ii h_z \sum_x \left( b_{0x}^z c_{1x} - b_{M+1,x}^z c_{Mx} \right).
\end{align}
After a Fourier transform, the Hamiltonian becomes
\begin{align}
    \HH &= \sum_{q \in [0,\pi]} \Psi_q^\dagger \mathscr{H}_q \Psi_q,
    &
    \Psi_q^\top &=
    \begin{pmatrix}
        b_{0,q}^z & c_{1,q} & c_{2,q} & \cdots & c_{M-1, q} & c_{M,q} & b_{M+1,q}^z
    \end{pmatrix},
\end{align}
where
\begin{equation}
    \mathscr{H}_q = 
    \begin{pmatrix}
        0 & -\ii h_z & 0
        \\
        \ii h_z & \alpha_q & \beta_q & \gamma_q
        \\
        0 & \beta_q^* & \alpha_q & \eta & \gamma_q
        \\
        & \gamma_q & \eta^* & \alpha_q & \beta_q & \gamma_q &
        \\
        & & \gamma_q & \beta_q^* & \alpha_q & \eta & \gamma_q
        \\
        & & & \ddots & \ddots & \ddots & \ddots &\ddots
        \\
        & & & & \gamma_q & \beta_q^* & \alpha_q & \eta & \gamma_q &
        \\
        & & & & & \gamma_q & \eta^* & \alpha_q & \beta_q & 0
        \\
        & & & & & & \gamma_q & \beta_q^* & \alpha_q & -\ii h_z
        \\
        & & & & & & & 0 & \ii h_z & 0
    \end{pmatrix}
    \quad
    \text{and}
    \quad
    \begin{cases}
        \alpha_q = -2\kappa \sin q, \\
        \beta_q = 2\ii K \cos (q/2), \\
        \gamma_q = 2\kappa \sin (q/2).
    \end{cases}
\end{equation}

The spectrum shown in Figs. 1(c) and 1(d) in the main text was obtained by diagonalizing the matrix $\mathscr{H}_q$ for a system with $M = 250$, $h_z = 0.15$, and $\kappa = h_z^3/\Delta_{2f}^2 \approx 0.049$, using the result $\Delta_{2f} = 0.263313$ from Ref.~\onlinecite{Panigrahi_PRB2023}.

%%%%%%%%%%%%%%%%%%%%%%%%%%%%%%%%%%%%%%%%%%%%%%%%%%%%%%%%%%%%%%%%%%%%%%%%%%%%%%%
\section{\normalsize Additional information on exact diagonalization}
%%%%%%%%%%%%%%%%%%%%%%%%%%%%%%%%%%%%%%%%%%%%%%%%%%%%%%%%%%%%%%%%%%%%%%%%%%%%%%%

Our ED calculations exploited translation symmetry to decompose the Hamiltonian into sectors of different momenta $\vec{k}$. If $T_1$ and $T_2$ denote the operators that implement translations by the vectors $\mathbf{a}_1 = (\sqrt{3},3)/2$ and $\mathbf{a}_2 = (0,3)$ shown in Fig.~1(a) of the main text, then each momentum sector is composed of states $\ket{\phi}$ satisfying $T_n \ket{\phi} = e^{\ii \mathbf{k} \cdot \mathbf{a}_n} \ket{\phi}$ for a different momentum $\mathbf{k} = \frac{\pi}{3\sqrt{3}} (2n_1-3n_2, \sqrt{3} n_2)$, with $n_1, n_2 \in \mathbb{Z}$.

\begin{figure}[t]
	\center
	\includegraphics[width=\textwidth]{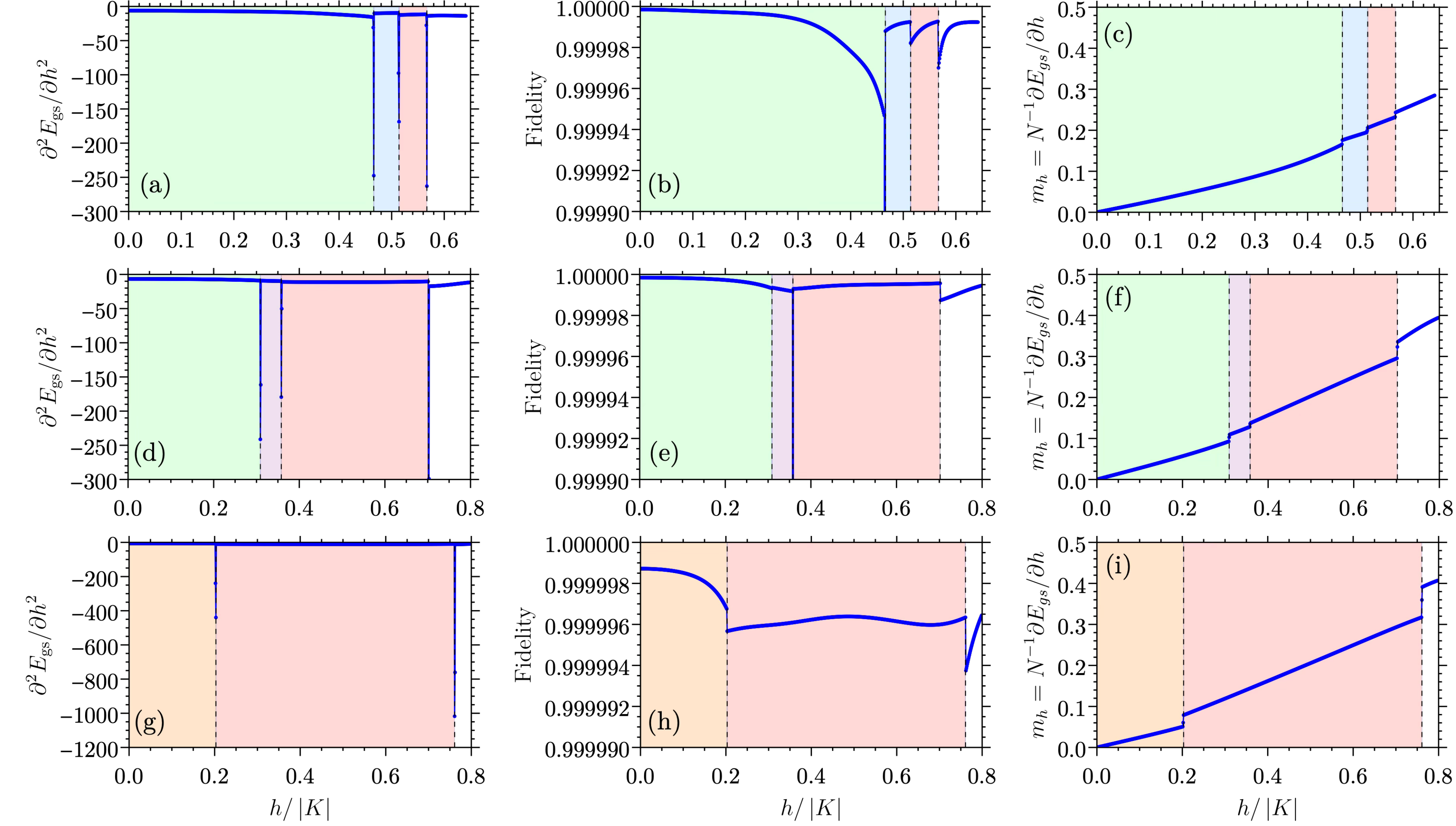} 
	\caption{ED results obtained for Model 1 at constant (a-c) $g = 0.02$, (d-h) $g = 0.10$ and (g-i) $g = 0.25$. Different regions in the plots are color-coded to corresponding ground state in Fig.~2(c) in the main text. The first column of plots shows to the second derivative of the ground-state energy with respect to $h$, $\partial^2 E_{\mathrm{gs}}/\partial h^2$, whereas the second and third columns depict the ground-state fidelity and the staggered magnetization per site. Note that the latter does not saturate in the field-polarized phase because a Néel-type product state where each spin points along the direction of the staggered field is not an eigenstate of the Hamiltonian.}
	\label{fig:eddata}
\end{figure}

We constructed the phase diagrams presented in the main text by performing scans at constant $g$ or $h$. Phase boundaries were determined by identifying the ground-state sector $\mathbf{k}$ at each point $\mathbf{x} = (g,h)$ and monitoring mainly two types of quantities: the second derivatives of the ground-state energy, $\partial^2 E_{\mathrm{gs}}/\partial g^2$ and $\partial^2 E_{\mathrm{gs}}/\partial h^2$, and the fidelity $\mathcal{F}(\mathbf{x}) = \abs{ \braket{\psi (\mathbf{x} + \delta \mathbf{x})}{\psi (\mathbf{x})} }$, which measures the overlap between ground-state wavefunctions at neighboring points, $\mathbf{x}$ and $\mathbf{x} + \delta \mathbf{x}$, with $\delta \mathbf{x} = (\delta g,0)$ or $(0,\delta h)$.
Representative results are displayed in Fig.~\ref{fig:eddata} to illustrate the procedure. Because of finite-size effects, continuous phase transitions tend to be associated with \emph{smooth} peaks in the second derivatives of the ground-state energy and the fidelity. By contrast, first-order transitions lead to sharp features that also manifest in observables such as the staggered magnetization per site
\begin{equation}
    m_h = \frac{1}{Nh} \sum_i \mathbf{h}_i \cdot \expval{\mathbf{S}_i},
    \label{eq:mh}
\end{equation}
depicted in the right panels.

Finally, we note that three different phases in the phase diagrams of the main text are realized within the nontrivial momentum sector $\mathbf{k} = \bm{K}$: (i) the narrow magenta phase extending from $h_{c1} \approx 0.4625 \abs{K_1}$ to $h_{c2} \approx 0.4775 \abs{K_1}$ for $g=0$; (ii) the nematic phase observed in Model 1; and (iii) the $X$ phase. This indicates that all such ground states pick up a nontrivial complex phase under translations.

%%%%%%%%%%%%%%%%%%%%%%%%%%%%%%%%%%%%%%%%%%%%%%%%%%%%%%%%%%%%%%%%%%%%%%%%%%%%%%%
\section{\normalsize Exact diagonalization in a tilted magnetic field}
%%%%%%%%%%%%%%%%%%%%%%%%%%%%%%%%%%%%%%%%%%%%%%%%%%%%%%%%%%%%%%%%%%%%%%%%%%%%%%%

To further substantiate the claim that the $X$ phase displays triple-$\mathbf{Q}$, and not single-$\mathbf{Q}$ magnetic order, we performed additional ED calculations in a tilted magnetic field
\begin{equation}
    \mathbf{h}_i = (-1)^i h \left( \cos \theta \, \unitvec{c} + \sin\theta\, \unitvec{b} \right),
\end{equation}
where $\unitvec{c} = \frac{1}{\sqrt{3}} (\unitvec{x} + \unitvec{y} + \unitvec{z})$ and $\unitvec{b} = \frac{1}{\sqrt{2}} (-\unitvec{x} + \unitvec{y})$ correspond to the usual [111] and [$\bar{1}$10] crystallographic directions. Thus, in an actual Kitaev material, $\unitvec{b}$ is parallel to one of the three types of bonds in the honeycomb plane. When $\theta = 0$ (as in the main text), the Hamiltonian in Eq.~(1) displays a discrete $C_3^*$ symmetry implemented by combining the 120° rotations in real space with the global spin rotations $(S_i^x, S_i^y, S_i^z) \mapsto (S_i^z, S_i^x, S_i^y)$. This symmetry enforces that any single-$\mathbf{Q}$ ground state, $\ket{\psi_1}$, has two degenerate partners, $\ket{\psi_2} = C_3^* \ket{\psi_1}$ and $\ket{\psi_3} = (C_3^*)^2 \ket{\psi_1}$, such that any linear combination $\ket{\psi} = \sum_{n=1}^3 c_n \ket{\psi_n}$ is also a ground state. It is because of this property that the structure factor from Fig.~3(a) cannot be conclusively ascribed to a triple-$\mathbf{Q}$ ground state.

\begin{figure}[t]
    \center
    \includegraphics[width=0.7\textwidth]{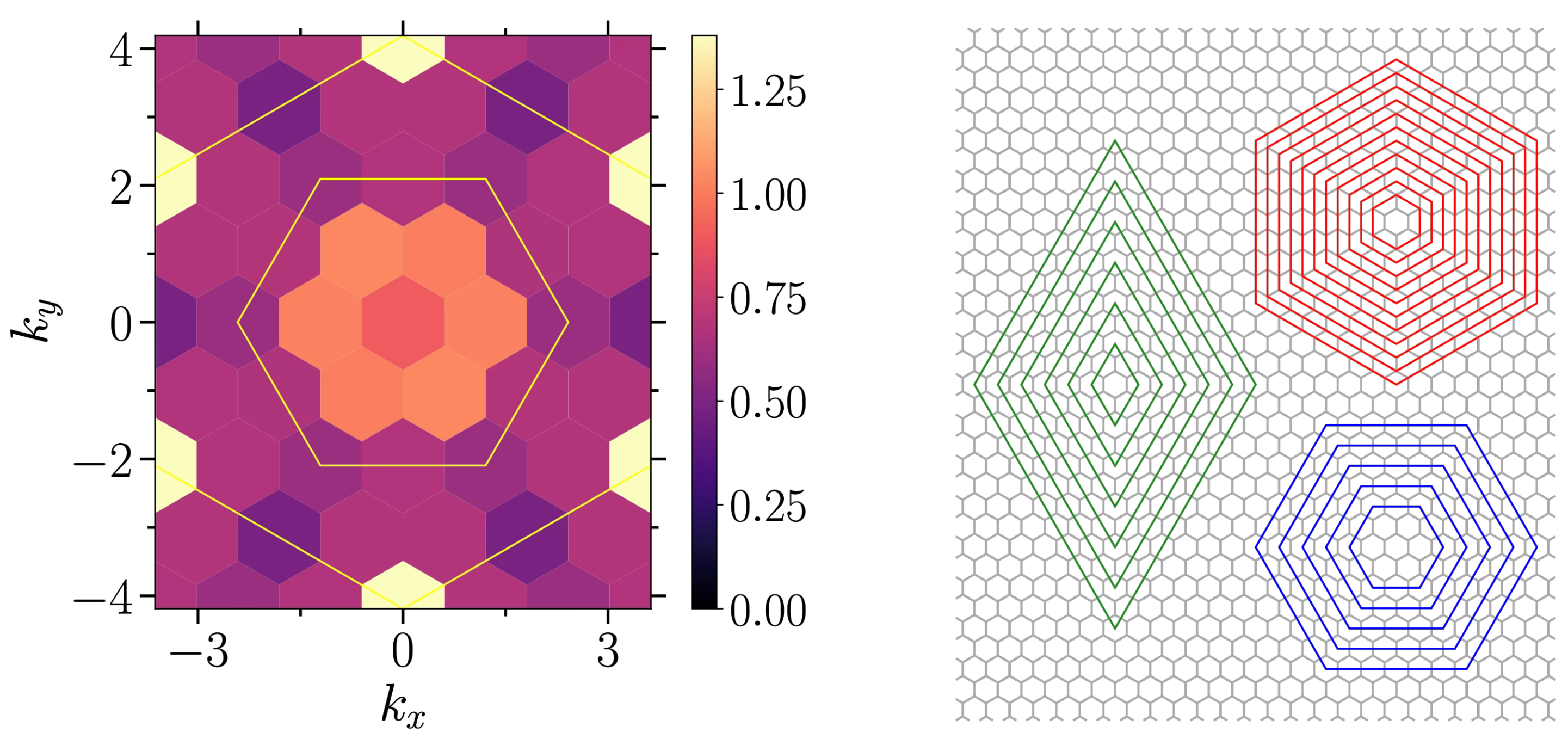} 
    \caption{Left: Static spin structure factor obtained deep within the $X$ phase for parameters $(g,h) = (0.75,0.55)$ and $\theta = 5^\circ$. The presence of three pairs of Bragg peaks in the absence of $C_3^*$ symmetry excludes the possibility that the $X$ phase shows single-$\mathbf{Q}$ magnetic order and, in agreement with our classical simulations, confirms the triple-$\mathbf{Q}$ scenario proposed in the main text. Right: Different clusters used in our iterative minimization study. The smallest blue cluster corresponds to the 24-site cluster utilized in our ED simulations.}
    \label{fig:ssf_tilt+clusters}
\end{figure}

Crucially, the inclusion of a tilt angle $\theta \ne 0$ breaks the above $C_3^*$ symmetry, and thereby lifts the degeneracy between $\ket{\psi_1}$, $\ket{\psi_2}$, and $\ket{\psi_3}$. As a result, the observation of three pairs of Braggs peaks in a spin structure factor is no longer compatible with a macroscopic superposition of single-$\mathbf{Q}$ domains and would confirm the existence of triple-$\mathbf{Q}$ order. In view of this argument, we computed the spin structure factor for the same parameters $(g,h) = (0.75, 0.55)$ used in Fig.~3(a), which place us deep in the $X$ phase, but now for a small tilt angle $\theta = 5^\circ$. As expected from our classical simulations, the results in Fig.~\ref{fig:ssf_tilt+clusters} corroborate the triple-$\mathbf{Q}$ nature of the magnetic order.

%%%%%%%%%%%%%%%%%%%%%%%%%%%%%%%%%%%%%%%%%%%%%%%%%%%%%%%%%%%%%%%%%%%%%%%%%%%%%%%
\section{\normalsize Iterative minimization}
%%%%%%%%%%%%%%%%%%%%%%%%%%%%%%%%%%%%%%%%%%%%%%%%%%%%%%%%%%%%%%%%%%%%%%%%%%%%%%%

Our classical simulations, whose results were discussed in the main text, made use of an iterative energy-minimization method related the algorithm presented in Ref.~\onlinecite{Walker_PRL1977}. Our algorithm is composed of the following steps:
\begin{enumerate}
    \item Given a finite-size cluster $\Lambda$, initialize the unit vectors $\left\{ \vec{S}_i \right\}$ in a random configuration.
    
    \item Compute the set of local fields
    \begin{equation}
        H_i^\alpha = h_i^\alpha - \sum_{j, \beta} J_{ij}^{\alpha\beta} S_j^\beta.
    \end{equation}
    Here, $J_{ij}^{\alpha \beta}$ is a matrix that includes all interactions between the spins at sites $i$ and $j$. It is defined by rewriting the Hamiltonian from Eqs.~(1) and (2) in the main text as $\mathcal{H} = \frac{1}{2}\sum_{i,j} \sum_{\alpha,\beta} S_i^\alpha J_{ij}^{\alpha\beta} S_j^\beta - \sum_i \vec{h}_i \cdot \vec{S}_i$.

    \item Generate a new spin configuration $\left\{ \vec{S}_i' \right\}$ defined by
    \begin{equation}
        \vec{S}'_i = \frac{1}{\mathcal{N}_i} \left[ r_i \unitvec{H}_i + (1-r_i) \vec{S}_i \right],
    \end{equation}
    where $\unitvec{H}_i = \vec{H}_i / \abs{\vec{H}_i}$, $r_i$ is a random number uniformly sampled from an interval $\left[ r_\mathrm{min}, r_\mathrm{max} \right]$ with $0 \le r_\mathrm{min} < r_\mathrm{max} \le 1$, and $\mathcal{N}_i$ is a normalization constant chosen to ensure that $\abs{\vec{S}_i'} = 1$. 

    \item Compute the deviations $\varepsilon_i = \abs{\vec{S}_i - \vec{S}_i'}$ and substitute $\left\{ \vec{S}_i' \right\} \to \left\{ \vec{S}_i \right\}$.

    \item Iterate steps 2-4 until the maximum deviation is less than a prescribed tolerance $\Delta$, i.e., $\max{(\varepsilon_i)} \le \Delta$.

    \item Repeat steps 1-5 for $N_\mathrm{init}$ different initial configurations and store the state $\left\{ \vec{S}_i^\Lambda \right\}$ with the lowest energy.
\end{enumerate}

In our simulations, we used simulation parameters $r_\mathrm{min} = 0.1$, $r_\mathrm{max} = 0.3$, $\Delta = 10^{-10}$, and $N_\mathrm{init} = 10^3$. Moreover, for each set of model parameters $g$ and $h$, we applied the iterative method to the 22 clusters depicted in Fig.~\ref{fig:ssf_tilt+clusters}, all of which were subject to periodic boundary conditions. In the end, we selected the state $\left\{ \vec{S}_i^\Lambda \right\}$ with the lowest energy per spin.

%%%%%%%%%%%%%%%%%%%%%%%%%%%%%%%%%%%%%%%%%%%%%%%%%%%%%%%%%%%%%%%%%%%%%%%%%%%%%%%
\section{\normalsize Computation of topological charges}
%%%%%%%%%%%%%%%%%%%%%%%%%%%%%%%%%%%%%%%%%%%%%%%%%%%%%%%%%%%%%%%%%%%%%%%%%%%%%%%

The topological charges $n_\mathrm{sk}^\mu$ ($\mu=A,B$) mentioned in the main text are computed as\cite{Berg_NPB1981,Diaz_PRL2019}
\begin{equation}
    n_\mathrm{sk}^\mu = \frac{1}{2\pi} \sum_{\uptr[0.2],\downtr[0.2]} 
    \arctan\left[ 
    \frac{\mathbf{S}_i \cdot \left( \mathbf{S}_j \times \mathbf{S}_k \right)}{1 + \mathbf{S}_i \cdot \mathbf{S}_j + \mathbf{S}_i \cdot \mathbf{S}_k + \mathbf{S}_j \cdot \mathbf{S}_k}
    \right],
    \label{eq:nsk}
\end{equation}
where the sum runs over all elementary plaquettes ($\uptr[0.3]$ or $\downtr[0.3]$) of sublattice $\mu$ in the magnetic unit cell. For any given plaquette, the sites $(i,j,k)$ are ordered in the same way, e.g., anticlockwise. Moreover, $\arctan$ refers to the principal branch of the arctangent function, whose image is $(-\pi/2,\pi/2]$. 
As a discretized version of the usual skyrmion number\cite{Nagaosa2013}, Eq.~\eqref{eq:nsk} is an integer that counts the number of times the spin configuration wraps the unit sphere.

%%%%%%%%%%%%%%%%%%%%%%%%%%%%%%%%%%%%%%%%%%%%%%%%%%%%%%%%%%%%%%%%%%%%%%%%%%%%%%%
\section{\normalsize Density functional theory calculations}
%%%%%%%%%%%%%%%%%%%%%%%%%%%%%%%%%%%%%%%%%%%%%%%%%%%%%%%%%%%%%%%%%%%%%%%%%%%%%%%

A bilayer was constructed from the experimental C2/m bulk structures for RuCl$_3$\cite{Cao_PRB2016} and MnPS$_3$ \cite{OUVRARD1985} including $\sim 20$ \AA\ of vacuum. The lattice constants were taken from MnPS$_3$ to mimic a monolayer of RuCl$_3$ atop a MnPS$_3$ bulk.  The atomic coordinates were then relaxed under a fully ferromagnetic order. For these calculations, we used the VASP code and the projector-augmented-wave (PAW) approach \cite{VASP,VASP-PAW} with the Perdew-Burke-Ernzerhof approximation of the generalized gradient approximation (GGA-PBE) as the exchange-correlation functional \cite{PBE}. An on-site Coulomb repulsion was added, following Dudarev's approach\cite{Dudarev1998}, with values of $U=3.0$ eV for Mn-$d$\cite{Amirabbasi2024} and $U=2.7$ eV for Ru-$d$ states\cite{Kaib2022}. The DFT-D3 van der Waals correction with zero-damping function was also included \cite{DFTD3}. 
Alternate van der Waals corrections were tested for atomic relaxations, still keeping the lattice constants fixed. The DFT-ulg method \cite{DFT-ulg}, the dDsC dispersion correction \cite{Steinmann2011_1,Steinmann2011_2}, and the Tkatchenko-Scheffler method \cite{Tkatchenko2009} were also explored and all metal-metal distances varied by less than 0.5\% with respect to the DFT-D3 case. An $8 \times 4 \times 1$ $\Gamma$ centered k-point grid was used for all these calculations with a 600 eV energy cutoff. Further, all calculations were done including spin-orbit coupling \cite{Kee_2016}.

\begin{figure}[t]
    \centering
    \includegraphics[width=0.6\linewidth]{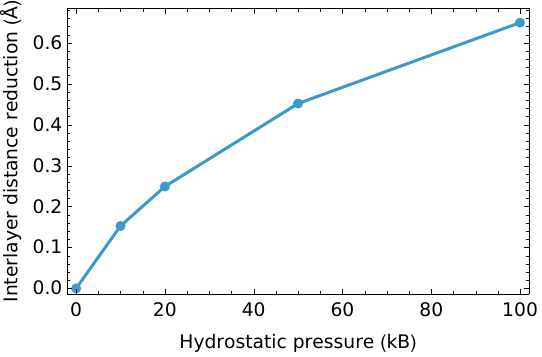}
    \caption{Reduction of bulk MnPS$_3$ interlayer distance under applied hydrostatic pressure.}
    \label{fig:pressure}
\end{figure}

Subsequently, the energies in a fully ferromagnetic state and a state consisting of ferromagnetic layers coupled anti-ferromagnetically out-of-plane were computed also using DFT+$U$+SOC. The characteristic coupling scale was obtained by calculating the energy difference between these two magnetic states.

Given that for the bare relaxed RuCl$_3$/MnPS$_3$ bilayer the derived magnetic coupling is too small to access the $X$ phase, we also performed pressure calculations to estimate the required interlayer distance (and related coupling) threshold to access this phase. To get this estimate, we proceed in two steps: To determine the effect of pressure on the interlayer distance, we relax bulk MnPS$_3$, allowing changes in both the internal coordinates and cell dimension, at fixed hydrostatic pressure. Fig.~\ref{fig:pressure} shows the interlayer distance reduction as a function of pressure. Applying a 5 GPa pressure we find a reduction in the interlayer distance as large as $0.45~ \mathrm{\AA}$. To determine the dependence of the effective field on the interlayer distance, we fix the structure within each layer and reduce the interlayer distance, moving the layers closer together. A reduction of just $0.1~ \mathrm{\AA}$ (which would correspond to a pressure of only 2 GPa) leads to an effective field $h \simeq 3.2$ meV, or values of $h/|K_1|\approx 0.64,0.42$ taking $K_1$ from models 1 and 2 in the main text, respectively. Comparing this to Fig.~2(c,d) in the main text, the derived effective field would be enough to realize the $X$ phase. This ultimately demonstrates that very achievable pressures would give access to the interlayer coupling needed to access the $X$ phase.

\bibliography{references}